\documentclass[11pt]{article}
\usepackage{graphicx,amsmath,amsfonts,amssymb}
\usepackage{color,multicol,multirow}
\def\sloppy{\tolerance=100000\hfuzz=\maxdimen\vfuzz=\maxdimen}
\allowdisplaybreaks
\def \beq  {\begin{equation}}
\def \eeq  {\end{equation}}
\def \beqar {\begin{eqnarray}}
\def \eeqar {\end{eqnarray}}
\def\bsp{\beq\begin{split}}
\def\esp{\end{split}\eeq}
\allowdisplaybreaks
\mathchardef\mhyphen="2D

\def\la {{\langle}}
\def\ra {{\rangle}}

\def\Tr {{\rm Tr}}
\def \tr {{\rm tr}}

\def\bigra{\bigr\rangle}

\def\del {\partial}
\def\bdel{\bar{\partial}}

\def\D {{\cal D}}

\def\half{\textstyle{1\over 2}}
\def\quarter {\textstyle{1\over 4}}

\begin{document}
\def \CMP {{Commun. Math. Phys.}}
\def \PRL {{Phys. Rev. Lett.}}
\def \PL {{Phys. Lett.}}
\def \NPBProc {{Nucl. Phys. B (Proc. Suppl.)}}
\def \NP {{Nucl. Phys.}}
\def \RMP {{Rev. Mod. Phys.}}
\def \JGP {{J. Geom. Phys.}}
\def \CQG {{Class. Quant. Grav.}}
\def \MPL {{Mod. Phys. Lett.}}
\def \IJMP {{ Int. J. Mod. Phys.}}
\def \JHEP {{JHEP}}
\def \PR {{Phys. Rev.}}
\begin{titlepage}
\null\vspace{-62pt} \pagestyle{empty}
\begin{center}
\rightline{CCNY-HEP-16/4}
\rightline{April 2016}
\vspace{1truein} {\Large\bfseries
The Geometry of Quantum Hall Effect: An Effective Action}\\
\vskip .15in
{\Large\bfseries for all Dimensions}\\
\vskip .1in
{\Large\bfseries ~}\\
{\large\sc Dimitra Karabali$^a$} and
 {\large\sc V.P. Nair$^b$}\\
\vskip .2in
{\itshape $^a$Department of Physics and Astronomy\\
Lehman College of the CUNY\\
Bronx, NY 10468}\\
\vskip .1in
{\itshape $^b$Physics Department\\
City College of the CUNY\\
New York, NY 10031}\\
\vskip .1in
\begin{tabular}{r l}
E-mail:&{\fontfamily{cmtt}\fontsize{11pt}{15pt}\selectfont dimitra.karabali@lehman.cuny.edu}\\
&{\fontfamily{cmtt}\fontsize{11pt}{15pt}\selectfont vpnair@ccny.cuny.edu}
\end{tabular}

\fontfamily{cmr}\fontsize{11pt}{15pt}\selectfont
\vspace{.8in}
\centerline{\large\bf Abstract}
\end{center}
We present a general formula for the topological part of the effective action for
integer quantum Hall systems in higher dimensions, including
fluctuations of the gauge field and metric around
background fields of a specified topological class.
The result is based on a procedure of integrating up from
the Dolbeault index density which applies for the degeneracies of Landau levels, 
combined with some input from the standard descent procedure
for anomalies.
Features of the topological action in (2+1), (4+1), (6+1) dimensions, including the contribution due to gravitational anomalies, are discussed in some detail.

\end{titlepage}

\pagestyle{plain} \setcounter{page}{2}
\setcounter{footnote}{0}
\setcounter{figure}{0}
\renewcommand\thefootnote{\mbox{\arabic{footnote}}}
\fontfamily{cmr}\fontsize{11pt}{16pt}\selectfont
\section{Introduction}

Quantum Hall effect has long been a fascinating phenomenon, 
both experimentally and theoretically \cite{books}.
 On the theoretical side, there has been the
immensely successful description based on wave functions.
It was also realized fairly early
 that the Chern-Simons action provides an effective description of many of the features
of quantum Hall effect.
For a quantum Hall droplet, i.e., for a finite system with boundary, the effective description must also include the action for a chiral boson theory. This is
needed to render the Chern-Simons action fully gauge-invariant and 
provides a description
for the edge excitations of the droplet.
The Chern-Simons action involves the electromagnetic vector potential and hence
pertains to the electromagnetic response of the QHE system; in fact, the effective action
incorporates the relevant transport coefficient, namely, the Hall conductivity.
Some of the other transport coefficients of interest, such as the Hall viscosity,
correspond to the response of the Hall system to perturbations of the background metric
\cite{viscosity, Read}.
Further,
considerations of the quantum Hall effect
on spaces of nontrivial topology can give insights into the physics of the problem,
even though
experimentally we may only be interested in spaces of trivial topology \cite{haldane, haldane2}.
As a result, the response of QHE systems to changes in the background metric,
captured via an effective action on spaces of different geometry and topology, 
has become the focus of many recent studies \cite{frohlich}-\cite{CR}.
The mathematical structures underlying quantum Hall effect have
also generated much interest in their own right, giving further impetus to such
studies.
 
Another branch of interesting generalizations of the QHE has been to higher dimensions \cite{HZ}-\cite{KN5}.
The Landau problem has been analyzed and  the wave functions and effective actions have been obtained for
a number of different spaces such as the four-sphere, complex projective spaces,
etc. In higher dimensions, 
the background gauge field can be Abelian or non-Abelian.
And as in the $(2+1)$-dimensional case, one can consider a bulk effective action
which captures the response to fluctuations of the gauge field.
The topological part of this effective action is a
generalization of the Chern-Simons action
to higher dimensions \cite{DK, VPN, KN5}.
Also in analogy with the lower dimensional case, one can consider a quantum Hall
droplet which would then allow for edge excitations, even when the background gauge field is
fixed, i.e., non-fluctuating.
The effective action for this has also been obtained in the case of integer filling fraction $\nu=1$; it is a generalization of the
WZW action \cite{KN2, KN3, KN4, poly1}.
Once fluctuations in the gauge field are also introduced, the calculated bulk and boundary actions 
were shown to be consistent with the mutual cancellation of anomalies
in the gauge symmetry \cite{DK}.\footnote{This is a generalization of the well known
similar structure in two dimensions \cite{wen2d}; the
cancellation of anomalies between the bulk and boundary terms is traceable to 
\cite{CalHarv}.}
The complete effective action captures the response of the system to
various gauge field perturbations and edge fluctuations of the droplet.

The natural question which arises from the juxtaposition of the two lines of development
outlined above would be: What is the effective action for QHE systems in higher dimensions,
including the response to gravitational fields?
This is the subject of the present paper. We will start with an index theorem for the
degeneracy of the quantum Hall states at various Landau levels on a K\"ahler manifold. 
This degeneracy is also the total charge (for the relevant gauge field)
of the fully occupied Landau level
and hence the response of the system to changes in the electrostatic-type
component (or time-component) of the corresponding vector potential.
The effective action can then be constructed, in essence, by integrating this
response with respect to the vector potential and making the result covariant.
We will use complex projective spaces to illustrate various aspects of these considerations,
but the result is general and applies to any K\"ahler manifold.

As mentioned above, many aspects of the
effective action in $(2+1)$-dimensions
with nontrivial geometry and topology have been considered
by several authors. In \cite{HS}, local Galilean invariance is used to
elucidate features of the effective action.
Effective actions, including gravitational contributions, are obtained in
\cite{AG1, Wiegmann} from microscopic dynamics. Geometric adiabatic transport has been
considered in \cite{KW1}-\cite{KW2}. 
In \cite{KW2}, the effective action is discussed from the
point of view of an index density.
There
are many points of concordance with
these papers,
when we specialize our general effective action
to $(2+1)$ dimensions; these will be referred to as the occasion arises.

This paper is organized as follows.  In the next section, we start with the Landau problem on
complex projective spaces and the degeneracy of the quantum Hall states at various Landau levels.
In section 3, we consider the degeneracy in terms of the relevant index theorem.
The index density can be identified as the charge density or the response to the 
time-component of 
an Abelian gauge field. We can then write down the gauge-field dependent
terms of the topological part of the effective action.
This is the action which would correspond to what is obtained by integrating out the fermions occupying the lowest Landau
levels. Generalization to higher Landau levels is taken up in section 4.
In section 5, we consider general gauge and gravitational fields and write down the effective
action for higher dimensions.
This action, given in (\ref{24d}), is the main result of this paper.
The gauge-field dependent
terms in this action are simplified in section 6 for
(4+1) dimensions, working out special cases in some detail;
section 7 addresses the same in (6+1) dimensions.
The contribution from the terms related to
the gravitational anomaly is considered in section 8, 
 with details worked out for (2+1), (4+1) and (6+1) dimensions explicitly.
A discussion section compares our results with the existing literature. There is 
a short appendix on some basic features and geometry of $\mathbb{CP}^k$ spaces.

For clarification we emphasize that, in this paper, we consider fully filled Landau levels (integer QHE) on manifolds without a boundary. What is obtained is the topological part of the bulk effective action. Fully filled Landau levels on manifolds with a boundary, droplets of finite size (with possible edge excitations) and the corresponding bulk and boundary actions are important issues. These will be left to future work. 

\section{ Landau levels and degeneracy}

As mentioned in the introduction, we will be concerned with the QHE and Landau levels
on spacetimes of the form
${\mathbb R} \times K$, where the spatial manifold $K$ is a complex manifold.
We will consider the case of $K$ being K\"ahler to begin with,
where the background magnetic field can be taken
as the K\"ahler two-form, up to a constant of proportionality, with the Hamiltonian
being proportional to the Laplace operator.
The states of the lowest Landau level (LLL) correspond to 
wave functions which satisfy a holomorphicity condition. More precisely, the wave functions
will be sections of an appropriate power of the line bundle
with the background field as the curvature.
In higher dimensions, non-Abelian background fields are possible, so
a slight generalization is needed.
Further, wave functions for the higher Landau levels can be considered
as the wave functions of the lowest Landau level of an equivalent
problem where the charged particles carry an appropriate amount of spin.
These statements are somewhat abstract and it is illuminating to
have an explicit construction. For most of the explicit examples,
we will consider the case of
$K$ being a complex projective space of complex dimension
$k$, i.e, $K = {\mathbb{CP}}^k$. So we start by setting up
the framework for QHE on ${\mathbb{CP}}^k$.

Since
${\mathbb{CP}}^k$ is the coset space ${SU(k+1) / U(k)}$,
the discussion is most easily carried out
following the group theoretic analysis given in \cite{KN1}-\cite{KN3}.
The group $SU(k+1)$ is the full group of continuous isometries of 
${\mathbb{CP}}^k$, with $U(k)$ as the isotropy group at each point.
Thus the representation of 
$U(k)$ for any field is the specification of its spin.
Further, the curvatures on the manifold take values in the 
Lie algebra of $U(k)$. In particular, they are constant in the tangent frame basis. (Explicit 
formulae for the curvatures on $\mathbb{CP}^k$ are given in the Appendix.)
It is then possible to consider additional ``constant" gauge background fields which
are proportional to these curvatures; more explicitly, we can have
an Abelian background corresponding to the $U(1)$ part of $U(k) \sim U(1) \times SU(k)$
and a non-Abelian background corresponding to the $SU(k)$ part.
This gives a well-posed Landau problem of
particle motion in a constant background field.

Let $t_A$, $A =1, 2, \cdots, k^2+2k$,  denote a basis of hermitian
$(k+1)\times (k+1)$-matrices viewed as the fundamental representation
of the Lie algebra of $SU(k+1)$.
We choose the normalization
by $\Tr \,(t_A t_B )=\half \delta_{AB}$.
The Lie algebra commutation rules, when needed, will be taken to be of the form
$[t_A, t_B] = i f_{ABC} \, t_C$, with structure constants
$f_{ABC}$.
The generators corresponding to the $SU(k)$ part of
$U(k) \subset SU(k+1)$ will be denoted by
$t_a$, $a =1, ~2, \cdots , ~ k^2 -1$ and the
generator for the $U(1)$
direction of the subgroup $U(k)$ will be denoted by 
$t_{k^2+2k}$. 

The Landau level wave functions 
can be considered
as functions on 
$SU(k+1)$ which have a specific transformation property under
the $U(k) \subset SU(k+1)$.
A basis of functions on the group $SU(k+1)$
is given by  the
matrices corresponding to the
group elements in a representation, or the so-called
Wigner $\cal{D}$-functions, which are defined as
\beq
\D^{(J)}_{\mathfrak{l}; \mathfrak{r}}(g) = \la J ,\mathfrak{l}\vert\, g\,\vert J, \mathfrak{r} \ra \label {5}
\eeq
where $\mathfrak{l}, ~\mathfrak{r} $ stand for two sets of quantum numbers specifying the 
states within the representation.
There is a natural left and right
action on an element $g\in SU(k+1)$, 
defined by
\beq
{\hat{L}}_A ~g = T_A ~g, \hskip 1in {\hat{R}}_A~ g = g~T_A
\label{6}
\eeq
where $T_A$ are the $SU(k+1)$ generators in the representation to which $g$ belongs.

There are $2k$ right generators of $SU(k+1)$ which are not in
the algebra of $U(k) \subset SU(k+1)$; 
these can be separated into $T_{+i}$, $i=1,2 \cdots ,k$, which are
of the raising type and $T_{-i}$ which are of the lowering
type. 
These generate translations while $U(k)$ generates rotations at a point.
We can thus define the covariant derivatives  on ${\mathbb{CP}}^k$
in terms of the right translation operators on $g$ as
\beq
D_{\pm i}  = i\,{{\hat R}_{\pm i} \over r}
\label{6a}
\eeq
where $r$ is a parameter with the dimensions of length.
(The volume of the manifold will be proportional to $r^{2k}$.)
Since the strength of the gauge field is given  by the commutator of covariant
derivatives, we can then specify the background magnetic field for our problem by
specifying the action of $U(k)$ on the wave functions; this is so because
the commutators of 
${\hat R}_{+i}$ and ${\hat R}_{-i}$ are in the
Lie algebra of $U(k)$.
The constant background field is given by the conditions
\beqar
{\hat R}_a ~\Psi^J_{m; \alpha} (g) &=&
(T_a)_{\alpha \beta} \Psi^J_{m; \beta} (g) \label{9a}\\
{\hat R}_{k^2 +2k} ~\Psi^J_{m; \alpha} (g) &=& - {n k\over \sqrt{2 k
(k+1)}}~\Psi^J_{m; \alpha} (g) \label{9b}
\eeqar
where $m=1,\cdots, {\rm dim}J$ counts the degeneracy of the Landau level.
Equation (\ref{9a}) shows that the wave functions $\Psi^J_{m; \alpha}$ transform,
under right
rotations, as a representation ${\tilde J}$
of $SU(k)$. Here
$(T_a)_{\alpha \beta}$ are the representation matrices for the
generators of
$SU(k)$ in the representation ${\tilde J}$, and
$n$ is an integer characterizing the Abelian part of the background field.
$\alpha ,\beta$ label states within the $SU(k)$ representation ${\tilde J}$
(which is itself
contained in the representation $J$ of $SU(k+1)$). The index $\alpha$ carried by the
wave functions $\Psi^J_{m; \alpha} (g)$
 is basically the gauge index. The wave functions are sections
of a $U(k)$ bundle on ${\mathbb{CP}}^k$.

The Hamiltonian $H$ for the Landau problem is proportional to the covariant Laplacian on 
${\mathbb{CP}}^k$; explicitly the action of $H$ on wave functions is given by
\beqar
H \, \Psi &=& - {1\over 4 m} (D_{+i} D_{-i} + D_{-i} D_{+i} ) \, \Psi\nonumber\\
&=&{1\over 2 m r^2} \left[ \hat{R}_{+i} \hat{R}_{-i} + {1\over 2} \left( i f_{-i, +i, a} \, \hat{R}_a +
i f_{-i, +i, k^2+2k} \, \hat{R}_{k^2 + 2k} \right) \right] \Psi
\nonumber\\
&=&{1\over 2 m r^2} \left[ \hat{R}_{+i} \hat{R}_{-i} +  {i \over 2} f_{-i, +i, a} \, T_a + 
{i \over 2} f_{-i, +i, k^2+2k} \, \left( - {n k \over \sqrt{2 k (k+1)}} \right) \right] \Psi
\label{9c}
\eeqar
We see that $H$ is proportional to 
$\sum_{i} {\hat R}_{+i} {\hat R}_{-i}$, apart from additive constants.
Thus the lowest Landau level should satisfy, in addition to the
requirements (\ref{9a}, \ref{9b}), the condition
\beq
\hat{R}_{-i} \, \Psi = 0
\label{9d}
\eeq
This is the holomorphicity condition on the lowest Landau level wave functions.
Thus the values of the background fields are specified or chosen by
(\ref{9a}), (\ref{9b}), which correspondingly set the choice of the
states $\vert J, \, \mathfrak{r} \ra \equiv \vert J, \, \alpha, w \ra$ in (\ref{5}), 
where $w = - n k/\sqrt{2 k (k+1)}$ is the eigenvalue of
$\hat{R}_{k^2 + 2 k}$,
and the lowest Landau level wave functions are holomorphic as in 
(\ref{9d}).

The degeneracy of the lowest Landau level for ${\mathbb{CP}}^k$ may be obtained
easily from group theory.
The relevant conditions are (\ref{9a}, \ref{9b}, \ref{9d}), or in terms of the state
$\vert J, \, \alpha, w \ra$,
\beqar
\hat{R}_{-i} \, \vert J, \, \alpha, w \ra &=& 0\label{10a}\\
\hat{R}_a \, \vert J, \, \alpha, w \ra = (T_a)_{\alpha \beta} \, \vert J, \, \beta, w \ra,
&\hskip .1in&
\hat{R}_{k^2 + 2k}  \, \vert J, \, \alpha, w \ra = - {n \, k \over \sqrt{2 k (k+1)}}
\, \vert J, \, \alpha, w \ra
\label{10b}
\eeqar
The state $\vert J, \, \alpha, w \ra$ must be a lowest weight state in the 
representation $J$ according to
(\ref{10a}). The weight vector of this state itself is specified by
(\ref{10b}). Thus the representation $J$ is fixed by
(\ref{10a}), (\ref{10b}), and its dimension will give the degeneracy. Explicit formulae for the degeneracy of the quantum Hall states on ${\mathbb{CP}}^k$ for arbitrary Landau levels have been derived in \cite{KN3}.
\section{The index theorem and the effective action for LLL}

There is another way to think about the degeneracy. The holomorphicity condition
(\ref{10a}) shows that the degeneracy, which is the number of normalizable solutions
to (\ref{10a}), may be obtained from the index theorem for the Dolbeault complex \cite{eguchi}. Since the wave functions respond to the background gauge fields as in (\ref{10b}), we need a version of the index theorem in the presence of
gauge fields; this is given by the twisted Dolbeault complex \cite{eguchi}.
This index theorem is given as
\beq
{\rm Index}( \bdel_V) = \int_K {\rm td}(T_cK) \wedge {\rm ch} (V)
\label{11}
\eeq
where ${\rm td}(T_cK)$ is the Todd class on the complex tangent space of $K$
and ${\rm ch}(V)$ is the Chern character of the vector bundle $V$ (given in terms of
traces of powers of the curvature of the vector bundle which is
also referred to as the field strength of the gauge field).
Explicitly,
the Todd class has the expansion
\beq
{\rm td} = 1 + {1\over 2} \, c_1 +{1\over 12} ( c_1^2 + c_2) + {1\over 24} c_1\, c_2
+ {1\over 720} ( - c_4 + c_1\,c_3 + 3 \, c_2^2 + 4\, c_1^2 \,c_2 - c_1^4) + \cdots
\label{25}
\eeq
where $c_i$ are the Chern classes. For any vector bundle with curvature ${\cal F}$, these
are given by
\beq
\det \left( 1 + {i \, {\cal F} \over 2 \pi} \,t\right) = \sum_i c_i \, t^i
\label{26}
\eeq
The Todd class may also be represented, via the splitting principle,
in terms of a generating function as
\beq
{\rm td} = \prod_i {x_i \over 1- e^{-x_i}}
\label{27}
\eeq
where $x_i$ represent the ``eigenvalues" of the curvature in a suitable canonical form
(diagonal or the canonical antisymmetric form for real antisymmetric $i\,{\cal F} $). 

The first few Chern classes for the complex tangent space can be explicitly written, using (\ref{26}), as
\beqar
c_1 (T_c K)& = & \Tr ~{iR \over 2\pi} \nonumber \\
c_2 (T_c K)& = & {1 \over 2} \Biggl[ \Bigl(\Tr {iR \over 2\pi}\Bigr)^2 - \Tr \Bigl({iR \over 2\pi}\Bigr)^2 \Biggr] \nonumber \\
c_3 (T_c K) &=& { 1 \over 3!} \Biggl[ \Bigl(\Tr {iR \over 2\pi}\Bigr)^3 - 3\, \Tr {iR \over 2\pi} \,\Tr \Bigl({iR \over 2\pi}\Bigr)^2 + 2 \,\Tr \Bigl({iR \over 2\pi}\Bigr)^3 \Biggr] \\
c_4 (T_cK) &=& { 1 \over 4!} \Biggl[ \Bigl(\Tr {iR \over 2\pi}\Bigr)^4 - 6 \Bigl(\Tr {iR \over 2\pi}\Bigr)^2 \,\Tr \Bigl({iR \over 2\pi}\Bigr)^2 + 8\, \Tr {iR \over 2\pi}\, \,\Tr \Bigl({iR \over 2\pi}\Bigr)^3\nonumber\\
&&\hskip .4in  +3\, \Tr \Bigl({iR \over 2\pi}\Bigr)^2 
\,\Tr \Bigl({iR \over 2\pi}\Bigr)^2 -6\,\Tr \Bigl({iR \over 2\pi}\Bigr)^4 \Biggr] \nonumber
\label{27a}
\eeqar
where $R$ is the curvature for $T_c K$. 
The Chern character, which is needed in (\ref{11}), is defined by
\beq
{\rm ch}(V) = \Tr \left( e^{i {\cal F} /2 \pi} \right) = {\rm dim}\,V + \Tr ~{{i {\cal F}} \over {2\pi}} + { 1 \over 2!} \Tr~ {{i{\cal F} \wedge i {\cal F}} \over {(2\pi)^2}} + \cdots
\label{28}
\eeq 
where ${\rm dim} V$ is the dimension of the bundle $V$.
(For now, ${\cal F}$ can be taken as $F$, the field strength due to the
external gauge field. Later, we will include the curvature of the spin bundle
in ${\cal F}$ as well.)

Since these classes are expressed in terms of the curvatures $R$ and $F$,
the index theorem gives a more general counting of states. The curvatures do not have to be the fixed, background values used in the group theoretic analysis, fluctuations of the metric and gauge fields
are automatically included.
For example, when $K$ is two-dimensional, the index reduces to
\beqar
{\rm Index}( \bdel_V) &=&  \int_{K} \left[ \Tr {{iF} \over {2\pi}}+ {\rm dim} V \,{ c_1 (T_c K) \over 2} \right] \nonumber \\
&=&  \int_{K} \Bigl[ {i F \over 2 \pi} + {i R \over 4 \pi}    \Bigr]
\label{12}
\eeqar
For ${\mathbb{CP}}^1 = SU(2)/U(1) \sim S^2$, only Abelian gauge fields are allowed, so ${\rm dim}(V)=1$. Further the corresponding background curvatures are (see Appendix)
\beq
\bar{F} = -i \,n\,\Omega, \hskip .2in \bar{R}\big\arrowvert_{T_cK}  = -i \, 2\, \Omega
\label{15}
\eeq
where $\Omega$ is the K\"ahler two-form on ${\mathbb{CP}}^1$. From now on we will denote the constant background fields by an overbar, while the unbarred quantities include fluctuations. Further, we take all connections and curvatures to be antihermitian.

For spinless  charged fields (i.e., ${\rm dim}\, V = 1$) and small fluctuations around the background fields given in (\ref{15})  the index works out to be 
\beq
{\rm Index}( \bdel_V) = (n+1) \int {\Omega \over 2\pi} = n+1
\label{15a}
\eeq
From the point of view of group theory, the conditions
(\ref{10a}), (\ref{10b}) tell us that the lowest Landau level
states form an $SU(2)$ representation with spin $j = {\half} n$, giving the degeneracy
$2 j +1 = n+1$, in agreement with (\ref{15a}).

The index theorem, however, gives the degeneracy for any general choice of curvatures, of which
(\ref{15}) are only a special case.
{\it We can therefore use the index density to construct an effective action
with an arbitrary metric and gauge field. This will be our basic strategy.}
(But $K$ should still remain a complex manifold for us to be able to use the
Dolbeault index.)

Continuing with the two-dimensional case, for a fully filled Landau level,
the number of states is identical to the total charge if we assign a unit charge to each
particle. Since the degeneracy of the lowest Landau level is given by the Dolbeault index, we can identify the corresponding index density with the charge density $J_0$ up to a total derivative term, i.e.,
\beq
J_0 =  {i F \over 2 \pi} ~+~ {i R \over 4 \pi} 
~+~ d\, M
\label{16}
\eeq
where $M$ is 1-form. (It will be a $(2k -1)$-form in general.) Further, the charge density $J_0$ is also the functional derivative of the effective action with respect to the time-component ($A_0$) of the $U(1)$ gauge field, 
\beq
{{\delta S_{\rm eff}} \over {\delta A_0} }= J_0
\eeq
Thus the effective action involving gauge fields can be obtained by ``integrating" the index density with respect to $A_0$, in other words, finding an $S_{\rm eff}$ such that
\beqar
\delta S_{{\rm eff}} &=&\int   (i \delta A_0 dx^0 ) \wedge  \left(  {i F \over 2 \pi} +{i R \over 4 \pi} \right) ~+~ i d\,(\delta A_0 dx^0) \wedge M
\nonumber\\
&=& \delta \Bigl[ {i^2 \over {4\pi}} \int A(F+R) \Bigr]
+ \delta {\tilde S}
\label{17}
\eeqar
(We use antihermitian components
for the gauge fields, including the time-component, which explains the additional
factors of $i$ in (\ref{17}).)
The effective action can thus be taken to be
\beq
S^{\rm LLL}_{3d} =  {i^2 \over {4\pi}} \int A(F+R)
+ S_{\rm grav} + {\tilde S}
\label{18}
\eeq
There is some explanation needed for the steps leading to (\ref{18}).
First of all, the Chern-Simons form involves terms with the time-derivatives
of the spatial components of the gauge potential, such as,
for example, $A\del_0 A$.
Our argument does not directly give these terms since there is no $A_0$ in such terms.
For the topological part of the action, our strategy is to
complete by covariance the result obtained from (\ref{17}) to arrive at (\ref{18}).
Secondly, there could be purely gravitational terms which cannot be determined from
(\ref{17}) since they are not $A_0$-dependent. The most important such terms
have to do with possible gravitational anomalies.
These will be taken up later; for the moment, $S_{\rm grav}$ in (\ref{18}) signifies such
terms.
Finally, since the charge density is specified as the index density only up to
an additive total derivative, as in (\ref{16}), there can be additional terms
of the form ${\tilde S} $ in (\ref{18}) whose variation gives
$i  (\delta A_0 dx^0 \, M)$. The term $d M$ in (\ref{16})
integrates to zero since we consider manifolds without boundary.
Thus the physics of a term like ${\tilde S}$ will involve dipole and higher moments
of the charge distribution of the filled Landau level.
Therefore, we can expect them to be subdominant in a 
derivative expansion of the effective action. Generically, they will also involve the metric
and hence would not qualify as topological terms. In (2+1) dimensions such terms have been derived under the assumption of local Galilean invariance \cite{HS} and explicitly calculated from the microscopic theory \cite{AG1, Wiegmann, KW2}.

We can now easily generalize these results to write down the topological bulk effective action describing the dynamics of the lowest Landau level with Abelian gauge fields for a complex space of arbitrary even spatial dimensions $2k$.
\beqar
S^{\rm LLL}_{2k+1} &=& \int \Biggl\{ \bigl[ 1 + {1\over 2} \, c_1 +{1\over 12} ( c_1^2 + c_2) + {1\over 24} c_1\, c_2 + \cdots \bigr]_{T_c K} \wedge \nonumber \\
&&\bigl[ i A + {i^2 \over {2(2\pi)}}  A F  + \cdots + {i^{l+1} \over {(l+1)!({2\pi})^l}}  A F^l  + \cdots \bigr] \Biggr\}_{2k+1}  + S_{\rm grav} + \tilde{S}
\label{18aa}
\eeqar
where the differential form of dimension $2k+1$ should be picked up in the integrand. Expression (\ref{18aa}) can be further generalized to include non-Abelian gauge fields. 

The general expression for the $(2k+1)$ dimensional Chern-Simons term (including Abelian and non-Abelian connections) can be written in the form
\beq
(CS)_{2k+1} (A)= {i \over k!} \int_0^1 d\tau\, \Tr \left[  A \, \left( {i F_\tau \over 2 \pi}\right)^k\right],
\hskip .3in
F_\tau =  \tau \, dA + \tau^2 \,A^2
\label{18b}
\eeq
One can check that its variation is of the form
\beq
\delta (CS)_{2k+1} =  {i\over k!}  \Tr \left[ \delta A \, \left( {i F \over 2 \pi}\right)^k\right]
\label{18a}
\eeq
Following similar reasoning as before, we can now write down the general bulk effective action for the lowest Landau level for any odd dimensional spacetime, for which the spatial part admits a complex structure. 
\beq
S^{\rm LLL}_{2k+1} = \int \Bigl[ {\rm td}(T_c K) \wedge  \sum_p (CS)_{2 p+1} (A)\Bigr]_{2 k+1}
~+~ S_{\rm grav}~+~\tilde{S}
\label{24}
\eeq
\section{ Effective actions for higher Landau levels}
So far we have considered the lowest Landau level.
The wave functions for the higher Landau levels do not satisfy a holomorphicity condition
like (\ref{10a}), so we cannot directly use the Dolbeault index. However, we
can use a simple trick to transform this to a lowest Landau problem for a charged particle carrying an appropriate amount of spin. For this, let us first consider the $s$-th Landau level
on ${\mathbb{CP}}^1$.
The wave functions are given by
\beq
\Psi_m(g) \sim \la J, m \vert\, g \, \vert J, -{\half} n\ra,
\hskip .2in J = {\half} n +s
\label{19}
\eeq
which has $R_3 \Psi = - {\half} n\, \Psi$ as required by
(\ref{10b}) but does not satisfy the holomorphicity condition (\ref{9d}). The states (\ref{19}) are however in the same representation
as
\beq
{\tilde \Psi}_m (g) \sim \la J, m \vert\, g \, \vert J, -{\half} n -s\ra
\label{20}
\eeq
for which the holomorphicity condition is satisfied, $R_{-} {\tilde \Psi} = 0$. 
We now consider a field $\phi$ which has $U(1)$ charge equal to
$1$ and which has $U(1)$ spin $s$.
Such a field couples to
the background field 
\beq
\bar{\cal F} =  -i \,(n +2s) \,\Omega
= \bar{F} + s \bar{R} = \bar{F} + \bar{\cal R}_s
\label{20a}
\eeq
This comes about because
the chosen background $U(1)$ gauge field is proportional to the
spin connection on ${\mathbb{CP}}^1$ (see Appendix). 
The lowest Landau level for this field will
obey a holomorphicity condition and, in fact, the wave functions
are given by
${\tilde \Psi}_m(g)$.
So the degeneracy for LLL of the field $\phi$ 
is the same as the degeneracy for
the $s$-th
Landau level for a spinless field with $U(1)$ gauge charge 1, which is the original field of interest.
Thus for the counting of states, we
can now use the Dolbeault index for the lowest Landau level for
$\phi$ (which is possible by virtue of the holomorphicity condition). 
Our strategy is to use this equality of degeneracies to
formulate the effective action in terms of the index density for
$\phi$.

The Dolbeault index is now written as
\beqar
{\rm Index}( \bdel_V) &=&  \int_{K} \left[ \Tr {{i(F+{\cal R}_s)} \over {2\pi}} + {\rm dim} V \,{ c_1 (T_c K) \over 2} \right] \nonumber \\
&=&  \int_{K} \Bigl[   {i F \over 2 \pi}+ (s+{\half} ) {i R \over 2 \pi}  \Bigr]
\label{20aa}
\eeqar 
For the particular values of $F,~R$ as in (\ref{20a}), this index counts correctly the degeneracy of the states in the $s$-th Landau level to be $n+1+2s$. 
In (\ref{20aa}), we can allow fluctuations in the fields,
so that
$F$ is the $U(1)$ magnetic field, $R$ is the curvature and ${\cal R}_s= s \, R$ is
the curvature of the spin bundle, all including fluctuations. 
(The choice of specific background values, as in (\ref{20a}), will be indicated
by barred quantities.)

Using (\ref{20aa}) and repeating the steps going from (\ref{12}) to (\ref{18}), we find
the bulk effective action for the filled $s$-th Landau level as
\beq
S^{(s)}_{3d} =  {i^2 \over {4\pi}} \int A\Bigr[ F + (2s+1)R \Bigl]
+ S_{\rm grav} + {\tilde S}
\label{22}
\eeq
The second term in (\ref{22}) arises from the coupling to gravity as discussed by
\cite{frohlich} and \cite{WZ} and is often referred to as the Wen-Zee term.
For us, $s = 0$ corresponds to the lowest
Landau level, so if we have $N$ filled Landau levels, the result would be
\beq
S = \sum_{s=0}^{N-1} S^{(s)}
\label{23}
\eeq

It is worth recapitulating the basic argument we have used.
Instead of dealing directly with the quantum Hall system in a higher Landau level, 
which we cannot do 
because of the lack of holomorphicity, we consider a mock system made of
particles with a suitably chosen value of spin, such that the lowest Landau
level of the mock system has wave functions in the same multiplet as
the original system at the required higher Landau level.
Since the degeneracies of the two systems are the same,
and since, at least for the (2+1) dimensional case,
the topological part of the response of the Hall system
depends only on the degeneracies or the index density, 
we can use the mock system to obtain the topological part of the effective action.
This is the basic strategy we are using.
\section{General fields and higher dimensions}
We can now extend these results to higher dimensional cases with $U(k)$ gauge fields and higher Landau levels, and gravitational fields,
guided by the discussion of the $\mathbb{CP}^1$ case.
For ${\mathbb{CP}}^k$, the field $\phi$, mentioned after (\ref{20}), couples to the constant background field 
\beq
\bar{\cal F} =-i \bigl( n\,\Omega\, \bold{1} + s \bar{R}^0 \bold{1} + \bar{R}^a T_a \bigr) = \bar{F} + \bar{\cal{R}}_s
\label{23aa}
\eeq
where $\bar{R}^0,~\bar{R}^a$ are the curvature components defined in (\ref{9A}) and $T_a,~\bold{1}$ are $U(k)$ matrices in the appropriate spin representation. 
With the addition of spin, the vector bundle whose Chern character enters the definition of the index in (\ref{11}) is the tensor product of the spin bundle and
the vector bundle for the internal gauge field. 
(By spin bundle, we do not necessarily mean the spinor bundle, but rather 
the bundle carrying a representation of the isotropy group of the manifold.
Also, for many examples, we will use the spin as a trick to get the action for higher
Landau levels, but we emphasize that this is not the only case of of interest.
One may also
consider the Hall effect for the lowest Landau level for particles of higher intrinsic spin.
Our considerations apply to such cases as well, with the suitable identification of the
various gauge fields and spin connections involved.)
Thus $V \rightarrow S \otimes V$.
The Chern character obviously splits into a product
${\rm ch}(S) \wedge {\rm ch}(V)$, 
\beq
{\rm ch} (S \otimes V) = \Tr \left(  e^{i ({\cal R}_s +F) /2 \pi} \right) 
= {\rm ch} (S) \wedge {\rm ch} (V)
\label{29a}
\eeq
In (\ref{29a}), ${\cal R}_s$ is the curvature $R$ in the representation appropriate to the chosen spin and the trace is
 over the spin module.
and $F$ is in the representation for the (gauge) charge rotations
of the field $\phi$. The spin connection which leads to
${\cal R}_s$ will be denoted by $\omega_s$ which will be valued in
the Lie algebra of $U(k)$.
 The connection for the bundle $S\otimes V$ is thus
 $\omega_s \otimes 1 + 1 \otimes A$ which we will often abbreviate as
 $\omega_s + A$.
 
The index theorem now becomes\footnote{
The zero modes of the $\bdel_V$ operator are also the lowest Landau
levels as in (\ref{9c}), (\ref{9d}). Thus the Dolbeault index is
what is relevant for us.
In \cite{dolan}, the zero modes of the Laplacian were analyzed by relating them to the zero
modes of the Dirac operator for a specific choice of the gauge potential being
proportional to the spin connection. For this choice, the index theorem for the Dirac operator 
can be written entirely in terms of the Chern classes for the gauge field. 
While this is adequate for evaluating the degeneracy, and response of the  system
to a limited variation in the fields which preserves the proportionality
of gauge potential and spin connection, we
are interested in considering arbitrary and independent fluctuations for the gauge and gravitational fields, so that an effective action for the response of the system to either or both can be obtained.
So a more general set-up is needed.}
\beq
{\rm Index}(\bdel_V) = \int_K {\rm td}(T_cK) \wedge {\rm ch} (S \otimes V) 
\label{23a}
\eeq
Upon taking the index density and following the steps which led from (\ref{16}) to (\ref{18}),
we can obtain an
effective action in $(2 k+1)$ dimensions. More directly, we can now introduce the Chern-Simons
forms by noting that
\beq
\delta \left[ \sum_p \Bigl\{ (CS)_{2 p +1} (\omega_s + A ) - (CS)_{2 p +1} (\omega_s) 
\Bigr\} \right] = \delta A  \wedge {\rm ch} (S\otimes V)
\label{23b}
\eeq
where $\delta A$ is the variation of the Abelian $U(1)$ component
of the gauge field. Since the term involving only $\omega_s$ in the expansion of
$(CS) (\omega_s +A)$ does not contribute in the variation,
we have subtracted it out on the left hand side of (\ref{23b}).
Such a term will contribute to the gravitational anomaly and will be discussed shortly.
The effective action can now be written as
\beqar
S^{(s)}_{2k+1} &=& \int \Bigl[ {\rm td}(T_c K) \wedge \sum_p \left[ (CS)_{2 p+1} (\omega_s + A) - (CS)_{2 p+1} (\omega_s) \right] \Bigr]_{2 k+1}
+ S_{\rm grav} + {\tilde S}\nonumber\\
&=& \int \Bigl[ {\rm td}(T_c K) \wedge \sum_p  (CS)_{2 p+1} (\omega_s + A)\Bigr]_{2 k+1}
- \int \Bigl[ {\rm td}(T_c K) \wedge \sum_p  (CS)_{2 p+1} (\omega_s )\Bigr]_{2 k+1}\nonumber\\
&& \hskip .5in
+ S_{\rm grav} + {\tilde S}
\label{24}
\eeqar
There are several observations to be made about this action. This action is in agreement with the well-known descent method used for anomalies \cite{anom}. Focusing first on just the gauge field dependent terms, and using
\beq
{1\over 2 \pi} d (CS)_{2 p +1} = {1 \over (p+1)!} \Tr \left(
{ i F \over 2 \pi}\right)^{p +1}
\label{24a}
\eeq
we see that the purely gauge field dependent part of the action (\ref{24}) may be considered as arising from the index density in
$(2 k+2) $ dimensions as
\beq
S =  2 \pi \int \Omega_{2 k+1} + \cdots ~, \hskip .5in
\bigl[{\rm Index ~Density}\bigr]_{2 k +2} = d \, \Omega_{2 k+1}
\label{24b}
\eeq
This relates our bottom-up approach of starting in $ 2 k$ spatial dimensions
to the descent approach used for the $(2+1)$-dimensional case in
\cite{KW2}. 
If we restrict the integration region in (\ref{24}), i.e, to a droplet,
the action (\ref{24}) will not be gauge-invariant; the lack of gauge invariance
is expressed as a boundary term.
This boundary term will be cancelled by the
anomaly of the $(2k-1, 1)$-dimensional theory of the edge excitations.
The anomaly of this $(2k-1, 1)$-dimensional theory is related to the index density
in $(2 k +2)$ dimensions in the standard descent procedure for anomalies.
The action (\ref{24}) is in accord with these expectations.

Such a descent method is known to apply to all anomalies, including the gravitational ones \cite{ag-witten} as well as the mixed gauge-gravity anomalies. The mixed terms are already apparent in (\ref{24}). To include the purely gravitational part and identify $S_{\rm grav}$ in (\ref{24}), we note that the gravitational anomaly can be obtained from the index density in
$(2 k+2)$ dimensions
from the appropriate terms in
${\rm td}(T_cK) \wedge {\rm ch} (S)$ \cite{ag-witten}. Using the definition of the Chern character in (\ref{28}), equation (\ref{24a}) and the fact that $d~[{\rm td}(T_cK)] =0$, we can write ${\rm td}(T_cK) \wedge {\rm ch} (S)$ as the exterior derivative of a $(2k+1)$-form as follows.
\beq
\left[ {\rm td}(T_cK) \wedge {\rm ch} (S)\right]_{2 k +2}
=  d\, \Omega_{2 k+1}^{\rm grav} + {1\over 2 \pi} \, d\, \Bigl[ {\rm td}(T_c K) \wedge \sum_p  (CS)_{2 p+1} (\omega_s )\Bigr]_{2k+1}
\label{24c}
\eeq
Here $d\,\Omega_{2 k+1}^{\rm grav} $ gives the $(2k+2)$-form in ${\rm td} (T_cK)$, namely $ [{\rm td} (T_cK)]_{2k+2} = d\, \Omega_{2 k+1}^{\rm grav}$.
Adding this term to (\ref{24}), we see that the effective action becomes
\beq
S^{(s)}_{ 2k+1} =
 \int \Bigl[ {\rm td}(T_c K) \wedge \sum_p  (CS)_{2 p+1} (\omega_s + A)\Bigr]_{2 k+1}
+ 2 \pi\int \Omega^{\rm grav}_{2k+1}
+ {\tilde S}
\label{24d}
\eeq
In this action we have gathered together the contributions from both gauge 
and gravitational fields. This result gives all the topological terms in the bulk 
effective action, encoding the
response of the system to gauge and gravitational fluctuations in arbitrary dimensions.

Finally, we note that in starting with the index density in $2 k$ dimensions and
interpreting it as the charge density for the Abelian field,
there is an ambiguity in writing down the effective action.
This is because several terms which only involve non-Abelian fields, such as, for example,
$(CS)_{2p+1}(A)$ where $A$ is in $\underline{SU(k)}$ do not contribute to the index and
hence the question of whether they are to be included in the effective action
or not is not settled by the index in $2k$ dimensions.
(The underlining of $SU(k)$ denotes the Lie algebra of the group.)
However, we know that there should be terms like
$(CS)(\omega_s)$ in the contribution due to the gravitational anomaly.
Further, it is the $S\otimes V$ bundle which is relevant and
hence there is some equivalence between the $A$'s and the $\omega$'s
once we restrict to the background fields. For this reason,
we should also have the purely non-Abelian $A$-dependent
terms in (\ref{24}) and (\ref{24d}).

To recapitulate, (\ref{24d}) gives the bulk effective action for the $s$-th higher Landau level for any
odd dimensional spacetime, for which the spatial part admits a complex structure, 
with $U(k)$ gauge fields.
(As mentioned, it can also be used for Hall effect in the lowest Landau level
for particles of arbitrary spin, with the suitable identification of
the fields.)
As always, for the topological terms,
the differential form of the appropriate dimension, namely $(2k +1)$,
must be picked out from the integrand in (\ref{24}) or (\ref{24d}); this is indicated by the
subscript.
While the topological terms follow from the index theorem, there can be non-universal, metric
dependent corrections which are indicated by ${\tilde S}$ in
(\ref{24}) and (\ref{24d}).

The effective action (\ref{24d}) is the main result of this paper.
Since it is still in rather cryptic form, we
will now consider working out the details of this action for some special cases and
for
certain choices of dimensions.
We will first consider the gauge-field dependent terms, since these are the ones relevant for
the counting of states. The terms which depend only on the 
gravitational fields will be taken up in section 8.

\section{$4+1$ dimensions: Gauge field dependent terms}

In the $(4+1)$ dimensional case, the part of the effective action depending on the
gauge fields reduces to 
\beqar
S_{\rm gauge} &=& \int \Biggl[  {{\rm dim} S\over 12} (c_1^2 + c_2)_ {(T_cK)}  +{1\over 2} c_1(T_cK) \wedge \Tr \left({{i {\cal R}_s} \over 2\pi}\right)+ { 1 \over 2} \Tr \left({i {\cal R}_s \over 2\pi}\wedge {i {\cal R}_s \over 2\pi}
\right) \Biggr] \wedge
(CS)_1(A) \nonumber \\
&& + \int \Biggl[ {{\rm dim} S\over 2} c_1(T_cK) + \Tr {i {\cal R}_s \over 2\pi} \Biggr] \wedge (CS)_3(A) + {\rm dim} S~\int (CS)_5 (A)\nonumber\\
&=& {{i^2} \over {(2\pi)^2}} \int \Biggl[ {{\rm dim}S \over 24 }  \bigl(  3\, (\Tr R)^2
- \Tr (R^2) \bigr)  + {1 \over 2}  (\Tr R) \wedge (\Tr {\cal R}_s ) + {1 \over 2} \Tr ({\cal R}_s)^2 \Biggr]
\wedge (CS)_1 (A) \nonumber\\
&& + { i \over 2 \pi} \int \Biggl[ {{{\rm dim} S} \over 2}\Tr R  + \Tr {\cal R}_s \Biggr] \wedge (CS)_3(A)
+ {\rm dim} S~\int (CS)_5 (A) 
\label{31a}
\eeqar
where, in the second expression, we have 
written out the characteristic classes explicitly.
The Chern-Simons terms are
\beq
\begin{split}
(CS)_1&= i \Tr (A), \hskip .3in (CS)_3 = {{i^2} \over 4\pi} \Tr \left[ A \, dA + {2\over 3} \,A^3\right]
\\
(CS)_5 &=   {{i ^3}\over {3! (2\pi)^2} }\Tr \left[
A \, dA\, dA + {3\over 2} A^3 \, dA + {3\over 5} A^5 \right]\\
\end{split}
\label{31b}
\eeq
and
\beq
R= -i \bigl[ R^0 {\bf 1} +  R^a t_a \bigr] ~~~~~~~~~~~~~~~ {\cal R}_s = -i \bigl[ s R^0 {\bf 1} +  R^a T_a\bigr] 
\label{31bb}
\eeq
with $t^a,~T^a$ being $SU(2)$ matrices in the fundamental and $j=s/2$ representation, respectively.
The action (\ref{31a}) is general, just restricting
(\ref{24d}) to $4+1$ dimensions.
The rest of this section will
be devoted to verifying that this is consistent with the 
expected degeneracies for various
special cases.

The index theorem which is associated with the action (\ref{31a}) is
\beqar
{\rm Index} (\bdel_V) &=&\!\! \!\int_K  {\rm dim} V \Biggl[  {{\rm dim} S \over 12}  (c_1^2 + c_2)_ {T_c K}  +{1\over 2} c_1(T_cK) \wedge \Tr \left({{i {\cal R}_s} \over 2\pi}\right) + { 1 \over 2} \Tr \left({i {\cal R}_s \over 2\pi}\wedge {i {\cal R}_s \over 2\pi}\right) \Biggr] \nonumber \\
&& \hskip .4in + \Biggl[ {{\rm dim} S\over 2} c_1(T_cK) + \Tr \left({i {\cal R}_s \over 2\pi} \right)\Biggr] \wedge \Tr {{iF} \over 2\pi} + {{\rm dim} S \over 2} \Tr \left({{iF} \over 2\pi}\wedge {{iF} \over 2\pi}\right) 
\label{32}
\eeqar
Our purpose will be to consider
 the index theorem for some special cases
 to show that the counting agrees
with what is obtained by explicit calculation of wave functions. This will justify the use of the
index density as the charge density for a $U(1)$ background and hence
justify the effective action (\ref{31a}).
\vskip .2in

\noindent{\underline  {1. ${\mathbb{CP}}^2$ with $U(1)$ gauge fields, lowest Landau level}}
\vskip .1in
As the first special case, we take $K$ to be ${\mathbb{CP}}^2
= SU(3)/U(2)$. In this case,
uniform background magnetic fields taking values in the Lie algebra of
$U(2)\sim SU(2) \times U(1)$ are possible.
As a first example then, we take the case of a magnetic field which is Abelian, corresponding to the
$U(1)$ subgroup of $U(2) \subset SU(3)$.
Further, we will consider a spinless field in the lowest Landau level, so that
$c_1 (S) = 0$, $c_2 (S) = 0$.
The vector bundle is one-dimensional, so ${\rm dim} V =1$.  Using the specific values of constant background fields for ${\mathbb{CP}}^2$ from (\ref{6A}-\ref{14A}), we find
\beq
\begin{split}
 \Tr \,{{i \bar{R}} \over 2\pi}  & = 3\, {\Omega \over 2 \pi}\\
\Tr \left({{i\bar{R}} \over 2\pi}\wedge {{i\bar{R}} \over 2\pi}\right) & = 3 \,\left({ \Omega \over 2\pi}\right)^2 \\
\int_{\mathbb{CP}^2} { 1 \over 12} ( c_1^2 + c_2 )\big\arrowvert_{T_cK}
&= 1
\end{split}
\label{37}
\eeq
The last line in (\ref{37}) holds, of course, even when fluctuations around the constant background values are included.
The background magnetic field is given by
\beq
\bar{F} = -i \,n\, \Omega
\label{38}
\eeq
The index theorem now gives
\beqar
{\rm Index}(\bdel_V) &=& {1 \over 12} \int  (c_1^2 + c_2)\big\arrowvert_{T_cK}
+ {1\over 2} \int c_1 (T_c K) \wedge  {iF \over 2\pi} ~+ ~
{1\over 2}\int  {{iF \wedge iF} \over (2\pi)^2}\nonumber\\
&=& 1+ {3n \over 2} + {n^2 \over 2} = {(n+1) (n+2) \over 2}
\label{39}
\eeqar

We can check this against the group theoretic derivation of the wave functions, which are 
proportional to $\la J, \mathfrak{l} \vert g \vert J, \mathfrak{r}\ra$.
A representation of $SU(3)$ may be taken to be of the $(p,q)$-type
corresponding to states of the form 
$ \big\arrowvert J, ~^{i_1 i_2\cdots i_q}_{j_1 j_2\cdots j_p} \bigra$
where each index (each of the $i$'s and the $j$'s)
can take values $1, 2, 3$. 
The upper indices transform as the ${\bf 3}^*$-representation, while the lower
ones correspond to the ${\bf 3}$-representation.
The states $\big\arrowvert J, ~^{i_1 i_2\cdots i_q}_{j_1 j_2\cdots j_p}\bigra$
are symmetric in all $p$ indices $i_1 ...i_p$, symmetric in all
$q$ indices $j_1 ...j_q$ and traceless.
The state $\vert J, \mathfrak{r}\ra$ must be a lowest weight state
with $\hat{R}_8 \vert J, \mathfrak{r}\ra = - (n/\sqrt{3}) \vert J, \mathfrak{r}\bigra$,
$\hat{R}_{-i }\vert J, \mathfrak{r}\ra = 0$. This identifies
the required representation as $(n,0)$ with
the state
$ \vert J, \mathfrak{r}\ra = \vert J, ~^{}_{33\cdots3}\bigra$ \cite{KN3}.
The dimension of the representation is thus ${\half} (n+1) (n+2)$, verifying
(\ref{39}).

\vskip .2in

\noindent{\underline  {2. ${\mathbb{CP}}^2$ with $U(1)$ gauge fields, $s$-th Landau level}}
\vskip .1in
Consider now the higher Landau levels, say, the $s$-th level,
taking $s = 0$ as the lowest level.
In this case, the required state is of the $(n+s,s)$ type with
$\vert J, \mathfrak{r}\ra = \big\arrowvert J,~^{3\cdots 3}_{33\cdots 3}\ra$.
This is not the lowest weight state; the lowest weight state in the same representation is
of the form $\big\arrowvert J,~^{i_1 i_2 \cdots i_s}_{33\cdots 3}\ra$ where the upper indices take
values $1, 2$. We can view this as the lowest Landau level
of a field with spin, specifically, with $SU(2)$ spin $j = {\half} s$ (hence
${\rm dim} S = 2 j +1 = s+1$),
$U(1)$ spin equal to $s$, and electric charge 1, coupling to the background field 
\beq
\bar {\cal F}= -i (n \Omega {\bf 1} + s \bar{R^0} {\bf 1} + \bar {R^a} T_a) = -i (\bar{F} {\bf 1} + \bar{\cal R}_s ) 
\label{39a}
\eeq
For such a field, $\big\arrowvert J,~^{i_1 i_2 \cdots i_s}_{33\cdots 3}\ra$
would be the lowest Landau level, satisfying the holomorphicity condition.
With these spin assignments, in addition to the Chern classes
in (\ref{37}), (\ref{38}), we find
\beq
\begin{split}
\Tr \left( {i \bar {\cal R}_s \over 2 \pi }\right) & = {{3s (s+1)}\over 2}\, {\Omega \over 2 \pi}\\
\Tr ~{{i\bar {\cal R}_s \wedge i \bar{\cal R}_s} \over {(2\pi)^2}}&= s (s+1) (2\,s-{\half}) \,\left( {\Omega \over 2\pi} \right)^2
\end{split}
\label{40}
\eeq
It is now easy to check that the index becomes
\beq
{\rm Index}(\bdel_V) = (s+1) \left[ {n^2 \over 2} + {3 n \over 2} (s+1) + ( s+1) ^2\right] = {{(s+1)(n+s+1)(n+2s+2)} \over 2}
\label{41}
\eeq
Group theoretically, the dimension of the $SU(3)$ $(n+s, s)$ representation
is the same as (\ref{41}) \cite{KN3}, justifying the use of the index density (\ref{32})
in constructing the effective action (\ref{31a}).

\vskip .2in

\noindent{\underline  {3. ${\mathbb{CP}}^2$ with non-Abelian gauge fields, lowest Landau level}}
\vskip .1in
As we mentioned before in the case of ${\mathbb{CP}}^k$, $k\ge 2$, there is a possibility of non-Abelian background gauge fields. In the case of ${\mathbb{CP}}^2$, the lowest Landau level states belong to a representation of $SU(3)$ with a lowest weight state which transforms nontrivially under $SU(2)$, as a representation $\tilde {J}$, as in (\ref{9a}). It was further shown in \cite{KN3}
that allowed ${\tilde J}$'s must correspond to
integer values of the spin ${\tilde j}$.

The background field is now purely of gauge nature (no coupling to spin connection), given by
\beq
\bar{\cal F}=-i \bigl( n \,\Omega\, {\bf 1} + \bar{R}^a \,T_a \bigr)
\label{41a}
\eeq
where $T_a$ are $(2 {\tilde j} +1) \times (2 {\tilde j} +1)$ matrices. Since there is no coupling to the spin connection, ${\cal R}_s$ can be set to zero in (\ref{31a}).
The index theorem (\ref{32}) now gives
\beqar
{\rm Index}(\bdel_V) &=& {{\rm dim}V \over 12} \int  (c_1^2 + c_2)\big\arrowvert_{T_cK}
+ {1\over 2} \int c_1 (T_c K) \wedge  \Tr\, {iF \over 2\pi} + 
{1\over 2} \int \Tr{{iF \wedge iF} \over (2\pi)^2}\nonumber\\
&=& (2{\tilde j} +1) \Bigl[ 1 + {3 \over 2} n + { 1 \over 2} n^2 - { 1 \over 2} {\tilde j} ({\tilde j} +1) \Bigr]
\nonumber\\
&=& {{(2 {\tilde j}  +1) (n+{\tilde j} +2) (n-{\tilde j} +1)} \over 2}
\label{41b}
\eeqar

This again agrees with the degeneracy of the lowest Landau level which is the dimension of the $SU(3)$ representation of the type $(p=n-{\tilde j} , q=2 {\tilde j} )$ \cite{KN3}.

\vskip .2in

\noindent{\underline  {4. ${\mathbb{CP}}^2$ with non-Abelian gauge fields and higher Landau levels}}
\vskip .1in
There are some intricacies when we consider a non-Abelian background gauge field and higher Landau levels.

The wave functions at the $s$-th Landau level form an $SU(3)$ representation
of the $(p,q)$ type with
$J = (p, q) = (n+ s - \tilde{j}, s+ 2 \tilde{j })$. They are of the form
$\la J, \mathfrak{l} \vert g \vert J, \mathfrak{r} \ra$, with
\beq
\vert J, \mathfrak{r} \ra = \big\arrowvert J,~^{33 \cdots 3; l_1 l_2 \cdots l_{2\tilde{j}}}_{33\cdots 3}\ra
\label{41d}
\eeq
There are $s$ upper 3's and $n + s - \tilde{j}$ lower 3's here. The $l$ indices indicate the non-Abelian gauge degrees of freedom.
This corresponds to a state with an eigenvalue of $\hat{R}^8$ equal to $- n /\sqrt{3}$ (as required)
 and transforming as the spin-$\tilde{j}$ representation of $SU(2)$.
(We also need $\tilde{j}$ to be an integer \cite{KN1}, \cite{KN3}; this is related to the fact that
 $\mathbb{CP}^2$ does not admit spinors.)
 The dimension of the representation is given by
 \beq
 {\rm dim} \,J = {1\over 2} ( n + 2 s + \tilde{j} +2) ( n + s - \tilde{j} +1) ( 2 \tilde{j} + s + 1)
 \label{41e}
 \eeq
As mentioned earlier these wave functions do not satisfy the holomorphicity condition. In order to be able to use the Dolbeault index as before, we convert this to a problem of lowest Landau level of a higher spin field. We consider the states $\tilde{\Psi} = \la J, \mathfrak{l} \vert g \vert J, \mathfrak{\hat{r}} \ra$ where
\beq
\vert J, \mathfrak{\hat{r}} \ra = \big\arrowvert J,~^{i_1i_2 \cdots i_s; l_1 l_2 \cdots l_{2\tilde{j}}}_{33\cdots 3}\ra
\label{41g}
\eeq
where there are $n + s - \tilde{j}$ lower 3's. The upper indices $i$ now indicate the spin and $l$ the gauge degrees of freedom. This state has $\hat{R}^8$ equal to $- n /\sqrt{3}$ (as required) and it is a lowest weight state. The representation it belongs to has dimension equal to (\ref{41e}) assuming that the indices $i,l$ in (\ref{41g}) are fully symmetrized. 

The corresponding field $\phi$ couples to the constant background field 
 \beq
 {\bar {\cal F}} = -i \left( n \Omega \, {\bf 1} +  s \bar{R}^0 \,  {\bf 1}+ 
 \bar{R}^a T_a \right)
 \label{41f}
 \eeq
where $T_a$ are $(2 j + s +1) \times ( 2 j + s +1)$ matrices. Fluctuations are then introduced as 
 \beq
  {\cal F} = -i \left(( n \Omega + \delta F)  {\bf 1} +  s (\bar{R}^0   + \delta R^0) {\bf 1} +
 (\bar{R}^a + \delta R^a)  T_a \right)
 \label{41h}
 \eeq
There is an ambiguity though of how to interpret the fluctuations $\delta R^a$. These can be thought of as either fluctuations of the non-Abelian gauge field or fluctuations of the non-Abelian spin curvature. In other words one can think of the field $\phi$ coupling to an Abelian gauge field and a $U(2)$ spin connection $(s, \tilde{j}+s/2)$ or coupling to a $U(2)$ non-Abelian gauge field and a $U(1)$ spin connection with spin $s$. Depending on the choice though, the effective action (\ref{31a}) will have a different field content. In particular the response to the metric will be different. On the other hand, the index (\ref{32}) evaluated for the background (\ref{41f}) will be exactly the same in both cases and equal to (\ref{41e}).

This ambiguity in constructing an effective action for a quantum Hall system with non-Abelian gauge fields at higher Landau levels has to do with the following. For the case of $\mathbb{CP}^2$, for example, recall that, for a field with spin which carries a nontrivial
$SU(2)$ gauge charge, the commutator of the covariant derivatives has the form
\beq
[ D_\mu, D_\nu ] \, \phi = -i \left( F_{\mu\nu} \, {\bf 1} + s R^0_{\mu\nu} \, {\bf 1} + F_{\mu\nu}^a \, t_a \otimes 1 +
R_{\mu\nu}^a\, 1 \otimes T_a   \right) \, \phi
\label{41c}
\eeq
where $F_{\mu \nu}$ is the $U(1)$ gauge field, $R^0_{\mu\nu}$ is the $U(1)$ spin curvature, $\{ t_a\}$ are in the representation of $\phi$ corresponding to the
gauge group action (say, $\tilde{j}$), $\{ T_a\}$ are in the representation
corresponding to the spin of $\phi$ (say, $s/2$ of $SU(2)$).
In the Landau problem, we choose the background value for the gauge field as
${\bar F}_{\mu\nu}^a = {\bar R}_{\mu\nu}^a $, where ${\bar R}_{\mu\nu}^a$ is the
standard curvature of $\mathbb{CP}^2$.
Thus, on the right hand side of
(\ref{41c}), we have the combination ${\bar R}_{\mu\nu}^a  ( t_a \otimes 1 + 1 \otimes T_a )$.
The group transformations generated separately by the $t_a$ and $T_a$ are not
important, only
the group action corresponding to the combination $(t_a \otimes 1 + 1 \otimes T_a )$ is relevant.
The wave functions which transform under the product of the two $SU(2)$'s corresponding to the gauge group and spin, namely, as $\tilde{j} \otimes s/2$,
can be reduced to irreducible components for the action of the combination
$( t_a \otimes 1 + 1 \otimes T_a )$. The $s$-th Landau level problem corresponds to a particular irreducible representation ($\tilde{j}+s/2$),
in the reduction of $\tilde{j} \otimes s/2$. (This corresponds to the full symmetrization of the indices $i_1\cdots i_s,l_1\cdots j_{2\tilde{j}}$ in (\ref{41g}).)

When we consider perturbations of the metric and the gauge field,
we then have two cases worthy of being distinguished.
If we consider perturbations which preserve the combination $( t_a \otimes 1 + 1 \otimes T_a  )$,
then the effective action can be obtained as the action 
with an Abelian gauge field and a curvature coupling
for a spin corresponding to the representation
$\tilde{j}+s/2$, or as the effective action with an Abelian gauge field and Abelian spin curvature and a non-Abelian gauge field of strength given by the representation $\tilde{j} + s/2$.
These two actions are not equivalent to each other although they give rise to the same index.
However, such perturbations are not the most general perturbations of the metric and the gauge field.
A general perturbation would consider independent values for
$F_{\mu\nu}^a = {\bar F}_{\mu\nu}^a + \delta F_{\mu\nu}^a$
and $R_{\mu\nu}^a = {\bar R}_{\mu\nu}^a + \delta R_{\mu\nu}^a$.
In this case, we can no longer classify wave functions under the combined $SU(2)$.
The perturbations couple different irreducible representations of the
combined $SU(2)$. 
In this case, we cannot sensibly consider integrating out
one Landau level (i.e. one irreducible representation in the reduction
of $\tilde{j} \otimes s/2$) to obtain an effective action.
One must consider all irreducible representations resulting from a given spin and
given gauge group representation. This corresponds to the case of lowest Landau level for a field with intrinsic spin and gauge degrees of freedom with a Hamiltonian proportional to the covariant $\bar{\partial}$ operator.  Such a field would couple to 
 \beq
  {\cal F} = -i \left(( n \Omega + \delta F)  {\bf 1} +  s (\bar{R}^0   + \delta R^0) {\bf 1} + (\bar{R}^a + \delta F^a)  t_a + (\bar{R}^a + \delta R^a)  T_a \right)
 \label{41k}
 \eeq
 where $t_a$ is in the $\tilde{j}$ and $T_a$ in the $s/2$ representation. We can now evaluate the index (\ref{41b}) for this background and we find it to be 
 \beq
 {\rm Index} = ( 2 j +1) ( s + 1) \left[ { n^2 \over 2} +{3n \over 2} (s+1) + (s+1)^2  - {1\over 2} j (j +1) \right]
\label{41n}
\eeq
As mentioned earlier when $\delta F^a=\delta R^a$, the states can be classified into multiplets corresponding to irreducible representations of the combined $SU(2)$ (of $t_a$ and $T_a$). These have spin values given by $J_i = \tilde{j} + {s \over 2}-i$, $i=1, \cdots, s$. The dimension for each of these multiplets is given by (\ref{41e}), where $\tilde{j} \rightarrow \tilde{j}-i$,
 \beq
 {\rm dim} \,J_i = {1\over 2} ( n + 2 s + \tilde{j} -i+2) ( n + s - \tilde{j} +i+1) ( 2 \tilde{j} -2i+ s + 1)
 \label{41j}
 \eeq
 It is straightforward to verify that summing over all these representations will produce the index in (\ref{41n}),
 \beqar
{\rm dim} &=& \sum_{i=0}^s (2 J_i +1) \left[
1 + {3 \over 2} \left( n + {3\over 2} s\right) + {1\over 2} \left( n + {3 \over 2} s\right)^2
- {1\over 2 } J_i (J_i +1) \right]\nonumber\\
&=& ( 2 j +1) ( s + 1) \left[ { n^2 \over 2} +{3n \over 2} (s+1) + (s+1)^2  - {1\over 2} j (j +1) \right]
\label{41s}
\eeqar

To briefly recapitulate the discussion in this subsection,
when we have a higher Landau level for, say, a spinless field, but
with a non-Abelian gauge field background,
we cannot directly use the index theorem as we do not have holomorphicity 
for the wave functions.
Translating the problem to a lowest Landau level problem for a field with spin,
we get fields of a certain spin as well as the non-Abelian charges.
The original Landau level of interest is one representation in the reduction of
the product of the spin representation and the gauge group representation of the field.
However, if we allow arbitrary fluctuations of the gauge field and the spin connection,
all representations in the reduction of the product mentioned above can occur.
Hence it is not possible to obtain an effective action for the original problem, i.e.,
just for the
higher Landau level of interest, by this method.
However, one can consider different but related physical situations.
One can write the
action for the field with spin and gauge charges (in the lowest Landau level),
from which we can obtain the response
of such a system to arbitrary independent variations of the gauge field and
the gravitational fields.
Or one can write an action for the
restricted case of identical fluctuations for the non-Abelian gauge field and the 
spin connection.
In this case, the response functions are also thus restricted.
 
\vskip .2in

\noindent{\underline  {5. $S^2 \times S^2$, arbitrary Landau levels}}

\vskip .1in
As another example, consider $K = S^2 \times S^2$. 
 In this case, 
\beq
R (T_cK) =  \left[ \begin{matrix} R&0\\ 0& {\tilde R} \\ \end{matrix} \right]
\label{42}
\eeq
where $R$ refers to the (antihermitian) curvature of the first $S^2$ and ${\tilde R}$ to the second.
Notice that $ \Tr (R\wedge R) =0$ for dimensional reasons,
so that $ c_2 (T_cK) = {\half} c_1^2$.
Considering Landau levels
$(s_1, s_2)$ corresponding to the two $S^2$'s, we have
\beqar
{i \bar{\cal R}_s \over 2 \pi} &=&  { s_1 \,\bar{R} + s_2 \bar{\tilde{R}} \over 2 \pi} = 2 { s_1 \,\Omega + s_2 {\tilde \Omega} \over 2 \pi} \nonumber \\
{i \bar{F} \over 2\pi} &=& {n_1 \Omega + n_2 {\tilde \Omega}\over 2\pi}
\label{44}
\eeqar
The index theorem can now be verified to be
\beq
{\rm Index} =  (n_1 + 2 s_1 +1 ) (n_2 + 2 s_2 + 1)
\label{45}
\eeq

In all these cases, namely the $\mathbb{CP}^2$ examples and the
$S^2 \times S^2$ example,
we see that the index density from (\ref{32}) does indeed reproduce the
correct counting of states and hence we can use it to construct the effective action,
which, of course, agrees with (\ref{31a}).

We close this section with a note about the normalization of the gauge fields.
We have taken the charge carried by the matter fields
for the Abelian gauge fields as unity, so that the number of states
(which is what the index theorem gives us) is equal to the
integral of the charge density.
But in writing the action, it is possible to use other normalizations.
For example, one might consider the Chern-Simons action
for the $U(k)$ gauge fields 
with the normalization of the $U(k)$
Lie algebra matrices fixed by their embedding in $SU(k+1)$.
While there is no particular motivation to do so, it may be useful
if one considers dimensional reduction of effective actions from a
higher dimension to a lower dimension.
The $U(1)$ charges in such a choice would not be unity, so the
normalization of the Chern-Simons term would be different from what is
given in (\ref{24}) or (\ref{31a}).
The appropriate normalization will follow from tracking the $U(1)$ charges of the relevant
matter fields of the Landau problem.

\section{6+1 dimensions: Gauge field dependent terms}

In $(6+1)$ dimensions, the part of the effective action which depends on the gauge fields is 
\beqar
S_{\rm gauge} &=& \int \Biggl[  {{\rm dim} S \over 24} c_1 c_2 +{{(c_1^2 + c_2)} \over 12}  \wedge \Tr {{i {\cal R}_s} \over 2\pi} +{c_1\over 2}  \wedge { 1 \over 2} \Tr \left({i {\cal R}_s \over 2\pi}\right)^2+ { 1 \over 3!} \Tr \left({i {\cal R}_s \over 2\pi}\right)^2 \Biggr] \wedge
(CS)_1(A) \nonumber \\
&+& \int \Biggl[  {{\rm dim}\, S\over 12} (c_1^2 + c_2)  +{1\over 2} c_1 \wedge \Tr {{i {\cal R}_s} \over 2\pi}+ { 1 \over 2} \Tr \left({i {\cal R}_s \over 2\pi}\right)^2 \Biggr] 
\wedge (CS)_3 (A) \nonumber \\
&+& \int  \Biggl[ {{\rm dim} S\over 2} c_1 + \Tr {i {\cal R}_s \over 2\pi} \Biggr] \wedge (CS)_5(A) + {\rm dim} S~\int (CS)_7 (A) + S_{\rm grav} + \tilde{S} 
\label{7d1}
\eeqar
Using the formulae for the Chern classes, this can be written more explicitly as 
\beqar
S_{\rm gauge} &=& {{i^3} \over {(2\pi)^3}} \int \Biggl[ {{\rm dim}S \over 48 }  \Bigl( (\Tr R)^3
- \Tr R \,\,\Tr (R^2) \Bigr)  + {1 \over 24} \left( 3(\Tr R)^2
-  \Tr (R^2) \right) \wedge (\Tr \,{\cal R}_s ) \nonumber \\
&&+ {1 \over 4} \Tr R \wedge \Tr ({\cal R}_s)^2  +{1 \over 3!} \Tr ({\cal R}_s)^3 \Biggr]
\wedge (CS)_1 (A) \nonumber\\
&+&{{i^2} \over {(2\pi)^2}} \int \Biggl[ {{\rm dim}S \over 24 }  \Bigl( 3(\Tr R)^2
-  \Tr (R^2) \Bigr) +{ 1 \over 2} \Tr R \wedge (\Tr {\cal R}_s ) + {1 \over 2} \Tr ({\cal R}_s)^2  \Biggr]
\wedge (CS)_3 (A) \nonumber\\
&&+ { i \over 2 \pi} \int \left[ {{{\rm dim} S} \over 2}\Tr R  + \Tr {\cal R}_s \right]\wedge (CS)_5(A)
+ {\rm dim} S~\int (CS)_7 (A) + S_{\rm grav} + \tilde{S}
\label{7d2}
\eeqar
where
\beq
R= -i \bigl[ R^0 {\bf 1} +  R^a t_a \bigr] ~~~~~~~~~~~~~~~ {\cal R}_s = -i \bigl[ s R^0 {\bf 1} +  R^a T_a \bigr] 
\label{7d3}
\eeq
with $t_a,~T_a$ being $SU(3)$ matrices in the fundamental and appropriate spin representation respectively.
The index associated with this action is 
\beqar
{\rm Index}(\bar{\partial}_V)_{6d} &=&\int {\rm dim}V \Biggl[  {{\rm dim} S \over 24} c_1 c_2 +{1 \over 12} (c_1^2 + c_2) \wedge \Tr {{i {\cal R}_s} \over 2\pi} +{1\over 2} c_1 \wedge { 1 \over 2} \Tr \left({i {\cal R}_s \over 2\pi}\right)^2 \nonumber\\
&&\hskip .8in +{ 1 \over 3!} \Tr \left({i {\cal R}_s \over 2\pi}\right)^2 \Biggr] ~+\nonumber\\
&&+ \int \Biggl[  {{\rm dim} S\over 12} (c_1^2 + c_2)  +{1\over 2} c_1 \wedge \Tr {{i {\cal R}_s} \over 2\pi}+ { 1 \over 2} \Tr ({i {\cal R}_s \over 2\pi})^2 \Biggr] 
\wedge \Tr{iF \over 2\pi}\nonumber\\
&&+ \int \Biggl[ {{\rm dim} S\over 2} c_1 + \Tr {i {\cal R}_s \over 2\pi} \Biggr] \wedge {1 \over 2} \Tr \left( {iF \over 2\pi}\right)^2 + { {{\rm dim} S} \over 3!} \int \Tr \left( {iF \over 2\pi}\right)^3
 \label{7d4}
\eeqar
As a check on the effective action, we can evaluate the index for a case
for which the degeneracy of the Landau level is known. 
Specifically, we will consider the special case corresponding to the QHE on ${\mathbb{CP}}^3$ with Abelian magnetic field at Landau level $s$ ($s=0$ corresponds to the lowest Landau level). The following relations are useful in evaluating the index:
\beqar
{\rm dim} S &=& {{(s+1)(s+2)} \over 2} \nonumber\\
\bar{F} = -i\,n \,\Omega ,  \hskip .3in \Tr{i \bar{R} \over 2\pi} & = & 4 {\Omega \over 2\pi}, \hskip .3in \Tr \left( {i \bar{R} \over 2\pi} \right)^2 = 4 \left( {\Omega \over 2\pi} \right)^2 \nonumber \\
\Tr { {i \bar {\cal R}_s} \over 2\pi} &=& {{s(s+1)(s+2)} \over 2} {4 \over 3} {\Omega \over 2\pi} \nonumber \\
 \Tr \left( { {i \bar{\cal R}_s} \over 2\pi} \right)^2 &=& {{(s+1)(s+2)} \over 2} {{(5s^2-1)} \over 3}  \left( {\Omega \over 2\pi} \right)^2 \nonumber \\
 \Tr \left( { {i \bar{\cal R}_s} \over 2\pi} \right)^3 & = &  {{(s+1)(s+2)} \over 2} \left(2s^3 -s^2 + {s \over 3}\right) \left( {\Omega \over 2\pi} \right)^3
 \label{7d5}
 \eeqar
 Using (\ref{7d5}) we find that the index can be written as
 \beq
{\rm Index} (\bar{\partial}_V)_{6d} = {{(s+1)(s+2)} \over 2} {{(n+2s+3)(n+s+1)(n+s+2)} \over 3!}
 \label{7d6}
 \eeq
 This is exactly the dimension of the $(n+s, s)$ $SU(4)$ representation which gives the degeneracy of the $s$-th Landau level for the abelian ${\mathbb{CP}}^3$ QH states \cite{KN3}.

\section{Full effective action including gravitational terms}

We now turn to the details of the
terms in the effective action related to the gravitational anomaly
in $(2+1)$, $(4+1)$ and $(6+1)$ dimensions. We will first consider these terms
separately, then combine them with the gauge field dependent terms discussed in the
previous sections to obtain the full effective action.
The result will, of course, correspond to the expansion of the full action
(\ref{24d}) for the appropriate dimension.

\subsection{ $(2+1)$ dimensional case}

In this case, we need those terms
in the index density 
in four dimensions which involve only the gravitational fields.
This is given by
\beq
{\rm Index~Density} (\bdel) = {{\rm dim} S \over 12}  (c_1^2 + c_2)_ {(T_cK)}  +{1\over 2} c_1(T_cK) \wedge \Tr {{i {\cal R}_s} \over 2\pi}+ { 1 \over 2} \Tr \left({i {\cal R}_s \over 2\pi}\right)^2 
\label{grav1}
\eeq
This follows from (\ref{32}) upon setting ${\rm dim} V = 1$ and $F= 0$.
Also, although we have spin $s$, since we are interested in two dimensions
eventually, we should keep in mind that
the fields have only one component; thus we can set ${\rm dim } S = 1$.
The various characteristic classes are,
\beqar
\left( c_1^2 + c_2 \right) (T_cK) &=& {i^2 \over {(2\pi)^2}} (d\omega)^2
\nonumber\\
\Tr ({i {\cal R}_s \over 2\pi})^2  &=& 
{i^2 \over {(2\pi)^2}}  s^2 (d\omega)^2\nonumber\\
c_1(T_cK)\wedge \Tr {{i {\cal R}_s} \over 2\pi} &=& {i^2 \over {(2\pi)^2}}s (d\omega)^2 \label{grav2}
\eeqar
where $\omega$ is the spin connection, $R=d \omega$.
The index density (\ref{grav1}) reduces to
\beq
{\rm Index~Density} (\bdel)  = {i^2 \over {2 (2\pi)^2}}  \left(s^2+s+{1 \over 6}\right)
d \left( \omega \, d \omega\right)
\label{grav3}
\eeq
The purely gravitational part of the topological effective action in (\ref{22})
for $(2+1)$ dimensions is thus given by
\beq
S_{\rm grav} = {i^2 \over 4\pi} \Bigl[ (s+{\half})^2 - {1\over 12} \Bigr]
\int  \omega \, d \omega
\label{grav4}
\eeq
Combining with the gauge-field part in (\ref{22}), the full topological bulk effective action for the $s$-th Landau level in $(2+1)$ dimensions is
\beqar
S_{3d}^{( s)} &=& {i^2 \over {4\pi}}\Bigr[  \int A \Bigl(dA +2(s+{1 \over 2})d\omega\Bigr) +   \Bigl( (s+{1 \over 2})^2 - {1\over 12} \Bigr)
\int  \omega \, d \omega \Bigl] \nonumber \\
&=& {i^2 \over {4\pi}} \int \Biggl\{\Bigl[A +(s+{1 \over 2})\,\omega\Bigr]d\Bigl[A +(s+{1 \over 2})\,\omega\Bigr] - {1 \over 12} \omega \,d\omega \Biggr\}
\label{3d}
\eeqar
This result agrees with \cite{AG1}, \cite{KW2}. (In our case $s=0$ corresponds to the lowest Landau level.)

\subsection{ $(4+1)$ dimensional case}

We now turn to the case of $(4+1)$ dimensions. 
The $6$-form index density for the gravitational fields
 is easily worked out from (\ref{23a}) as
\beqar
{\rm Index ~Density}(\bdel) &=&
{\rm dim} V \left[
{{\rm dim}S \over 24} c_1 c_2 + {c_1^2 + c_2 \over 12} \, {\rm ch}_1(S)
+ {c_1\over 2} \, {\rm ch}_2 (S)  + {\rm ch}_3 (S) 
\right]\nonumber\\
{\rm ch}_k (S) &=& {1\over k!} \Tr \left( {i {\cal R}_s \over 2 \pi}\right)^k
\label{grav5}
\eeqar
For the four-dimensional $K$, the holonomy group being $U(2)$, the curvatures take
values in the Lie algebra of $U(2)$, so $R$ is of the form
\beq
R (T_c K) = -i (R^0 {\bf 1} + t_a \, R^a)
\equiv d\,\omega^0 +{\tilde R} 
\label{grav6}
\eeq
where $t_a$ are the $SU(2)$ generators in the fundamental representation, ${\bf 1}$ is the $2 \times 2$ identity matrix, $\omega^0$ is the $U(1)$ connection and $\tilde{R}$ is the $SU(2)$ curvature. The curvature for the spin bundle is
\beq
{\cal R}_s = -i( s   R^0 {\bf 1}+  R^a T_a)
\label{grav7}
\eeq
where $T_a$ is in some spin $j$ representation of $SU(2)$ and 
${\bf 1}$ is the $(2 j+1)\times (2j +1)$ identity matrix. For generality we can keep $s$, $j$ independent from each other. In the particular case where we want to write down the effective action for spinless charged particles for the $s$-th Landau level of $K={\mathbb{CP}}^2$, we need to identify 
$j= {\half} s$.

The index density works out to be
\beqar
{\rm Index~Density}(\bdel )
&=& {i^3 \over {(2\pi)^3}} {( {\rm dim} V ) (2 j+1) (s+1)  \over 12}
\biggl[ (2 s^2 + 4 s + 1 ) (d\omega^0)^3 \nonumber\\
&+&
{ 8 j (j+1) - 1\over 4} ~ {d\omega^0} \wedge (-iR^a) \wedge (-i R^a) \biggr]
\label{grav8}
\eeqar
We then identify the gravitational contribution to the effective action as
\beqar
S_{\rm grav} &=& {i^3 \over {(2\pi)^2}} ( {\rm dim} V ) (2 j+1) (s+1)
\biggl[ ~{1 \over 6}\Bigl( (s+1)^2 -{1 \over 2}\Bigr) \int \omega^0\, (d \omega^0 )^2\nonumber\\
&&\hskip 1.6in
+ \Bigl( {1 \over 3} j (j+1) - {1\over 24} \Bigr)~ \int \omega^0 \, \Tr ( {\tilde R} \wedge {\tilde R})\biggr]
\label{grav9}
\eeqar
where $\tilde{R}$ indicates the $SU(2)$ curvature and $\Tr (\tilde{R} \wedge \tilde{R}) = \half (-iR^a)\wedge (-i R^a)$.
There are alternate ways to write this. For example, in the last term, we can 
replace the integral by a partial integration as
\beq
\int \omega^0 \, \Tr ({\tilde R} \wedge {\tilde R}) = \int d\omega^0 \, \Tr \left( {\tilde \omega}
d {\tilde \omega} + {2\over 3} {\tilde \omega}^3\right)
\label{grav10}
\eeq
where $\omega^0$, ${\tilde \omega}$ are the connections for the $U(1)$ and $SU(2)$ curvatures.
Since we are considering manifolds without boundary, these different forms are
equivalent.
(The boundary at the limits of the time-integration are not
null, and so these different ways would correspond to
different ways of writing the symplectic form, if one proposes to set up a
Hamiltonian version of the effective
action.)

One can now combine (\ref{31a}) and (\ref{grav9}) to write down the full topological action in $(4+1)$ dimensions. (This is, of course, equivalent to
the $(4+1)$-form from the action (\ref{24d}).)
For simplicity, we will only consider an Abelian gauge field now.
The gauge part of the action in (\ref{31a}) can then be written as
\beqar
S_{\rm gauge} &=& {i^3  (2j+1)\over {(2\pi)^2}} \int \Biggl\{ {1 \over 2} \Bigl[ (s+1)^2 - {1 \over 6}\Bigr]\, A \,(d\omega^0)^2 + \Bigl[{1 \over 3} j(j+1) - {1 \over 24}\Bigr] \,A  \,\Tr (\tilde{R} \wedge \tilde{R}) \nonumber \\
&&+\, {{(s+1)} \over 2} ~A\, dA\, d\omega^0 + { 1 \over 3!} \,A\, (dA)^2 \Biggr\} \label{CP21} \\
&=& {i^3(2j+1) \over {(2\pi)^2}}  \int \Biggl\{ { 1 \over 3!} \Bigl(A+(s+1)\omega^0\Bigr) \Bigl[d(A+(s+1)\omega^0)\Bigr]^2 -{ (s+1)^3 \over 3!} \omega^0 (d\omega^0)^2 \nonumber \\ 
&& -{ 1 \over 12} A (d\omega^0)^2 
+\Bigl( {1 \over 3} j (j+1) - {1\over 24} \Bigr) ~A \, \Tr ( {\tilde R} \wedge {\tilde R}) \Biggr\}
\label{CP2}
\eeqar
The first four terms in (\ref{CP21}) constitute the analog of the Wen-Zee term in (4+1) dimensions while the last term is the gauge Chern-Simons term. 
Combining (\ref{CP2}) and (\ref{grav9}) and setting ${\rm dim} V=1$ we find the full topological action
\beqar
S_{5d}^{(s)} &=&{i^3 (2j+1)\over {(2\pi)^2}} \int \Biggl\{ { 1 \over 3!} \Bigl(A+(s+1)\omega^0\Bigr) \Bigl[d\Bigl(A+(s+1)\omega^0\Bigr)\Bigr]^2 \nonumber \\
&&-{ 1 \over 12} \Bigl(A+(s+1)\omega^0\Bigr) \Biggl[  (d\omega^0)^2 -\Bigl[ (4j(j+1) - {1\over 2} \Bigr]\Tr ( {\tilde R} \wedge {\tilde R}) \Biggr] \Biggr\}
\label{CP22}
\eeqar
Further setting $j=s/2$ in (\ref{CP22}) will give the bulk topological action for the $s$-th Landau level QHE on ${\mathbb{CP}}^2$ with Abelian magnetic fields. 
Notice that an interesting effect of the gravitational interaction is to replace $A \rightarrow A+(s+1) \,\omega^0$ in (\ref{CP22}). The analog effect in the case of ${\mathbb{CP}}^1$ was $A \rightarrow A+(s + \half) \omega$ as in (\ref{3d}).

\subsection {$(6+1)$ dimensional case}

In $(6+1)$ dimensions we need to evaluate the $8$-form index density. Again, for simplicity we will consider the case of Abelian magnetic fields (${\rm dim} V=1$); we will also
consider only the case of spin zero fields, $s=0,~{\cal R}_s=0, {\rm dim}S=1$ (lowest Landau level). The corresponding index density involving gravitational fields is
\beq
{\rm Index~Density} (\bdel) = {1 \over 720}  (-c_4+c_1c_3 + 3c_2^2 + 4 c_1^2 c_2 -c_1^4)
\label{g6}
\eeq
Using the expressions for the characteristic classes in (\ref{27a}) we find 
\beq
{\rm Index~Density} (\bdel) = {1 \over 720}  \Biggl\{  {15 \over 8} \Bigl(\Tr {iR \over 2\pi}\Bigr)^4 - {15 \over 4} \Bigl(\Tr {iR \over 2\pi}\Bigr)^2 \,\Tr \Bigl({iR \over 2\pi}\Bigr)^2 +{ 5 \over 8} \, \Biggl[\Tr \Bigl({iR \over 2\pi}\Bigr)^2 \Biggr]^2 +{1 \over 4} \,\Tr \Bigl({iR \over 2\pi}\Bigr)^4 \Biggr\}
\label{g7}
\eeq
where 
\beq
R  = -i (R^0 {\bf 1} + R^a t_a)
\equiv d\,\omega^0 +{\tilde R} 
\label{g8}
\eeq
where $t_a$ are the $SU(3)$ generators in the fundamental representation, ${\bf 1}$ is the $3 \times 3$ identity matrix, $\omega^0$ is the $U(1)$ spin connection and $\tilde{R}$ is the $SU(3)$ curvature. 

From (\ref{g8}) we find that the purely gravitational contribution to the topological action in (6+1) dimensions is
\beq
S_{\rm grav} = {1 \over (2\pi)^3} {1 \over 720} \int \Biggl\{ 57 \,\omega^0 d\omega^0 \Bigl[ (d\omega^0)^2- {1 \over 2} \Tr (\tilde{R} \wedge \tilde{R}) \Bigr] + \omega^0 \,\Tr (\tilde{R}\wedge \tilde{R}\wedge \tilde{R}) \Biggr\} + { 1 \over 120} \int (CS)_7(\tilde{\omega}) 
\label{g9}
\eeq
where $\tilde{\omega}$ is the $SU(3)$ spin connection and
\beq
CS_{7} (\tilde{\omega})=  {1\over {4! (2\pi)^3}}  \Tr \left[ \tilde{\omega} (d\tilde{\omega})^3 + {12 \over 5} \,\tilde{\omega}^3 (d\tilde{\omega})^2 +2 \,\tilde{\omega}^5 (d\tilde{\omega}) +{4 \over 7} \,\tilde{\omega}^7 \right]
\label{g10}
\eeq
The gauge contribution to the topological action (\ref{7d2}) for an Abelian magnetic field and spin zero fields (LLL) is
\beqar
S_{\rm gauge} &=& { 1 \over (2\pi)^3} \int \Biggl\{{ 1 \over 4!} \left(A + {3 \over 2} \omega^0\right) \left[ d\left( A + {3 \over 2} \omega^0\right)\right]^3\nonumber\\
&&\hskip .7in - {1 \over 16} \left(A + {3 \over 2} \omega^0\right)d\left(A + {3 \over 2} \omega^0\right) \left[ (d\omega^0)^2 + {1 \over 3} \Tr (\tilde{R} \wedge \tilde{R}) \right] \nonumber \\
&& \hskip .7in -{ 9 \over 128} \omega^0d\omega^0 \left[  (d\omega^0)^2 -{2 \over 3} \Tr (\tilde{R} \wedge \tilde{R}) \right] \Biggr\}
\label{g11}
\eeqar
Adding (\ref{g9}) and (\ref{g11}) we get the bulk topological action for the lowest Landau level of 
${\mathbb{CP}}^3$ with Abelian magnetic fields. The full action is
\beqar
S_{7d}^{\rm LLL} &=& { 1 \over (2\pi)^3} \int \Biggl\{ { 1 \over 4!} \left(A + {3 \over 2} \omega^0\right) \left[ d\left( A + {3 \over 2} \omega^0\right)\right]^3\nonumber\\
&&\hskip .7in - {1 \over 16} \left(A + {3 \over 2} \omega^0\right)d\left(A + {3 \over 2} \omega^0\right) \left[ (d\omega^0)^2 + {1 \over 3} \Tr (\tilde{R} \wedge \tilde{R}) \right] \nonumber \\
&&\hskip .7in +{ 1 \over 1920} \omega^0d\omega^0 \left[ 17 (d\omega^0)^2 + 14 \Tr (\tilde{R} \wedge \tilde{R}) \right] + {1 \over 720} \omega^0 \Tr (\tilde{R} \wedge \tilde{R}\wedge \tilde{R}) \Biggr\} 
\nonumber\\
&&+ {1 \over 120} \int (CS)_7 (\tilde{\omega})
\label{g12}
\eeqar
Again, this corresponds to the appropriate simplification
of the general action (\ref{24d}).
In (\ref{g12}) we see again the shift $A \rightarrow A+ {3 \over 2} \omega^0$ in the presence of gravitational interactions.

\subsection {Comments}

It is worth pointing out 
a couple of interesting features of the gravitational contributions.
First of all, we notice that the $U(1)$ part of the spin connection
combines with
the $U(1)$ gauge field, as $A+ ( s+ {\half } k ) \omega^0$, for 
$\mathbb{CP}^k$. This is explicitly seen for the cases we have considered, namely, for $k=1, \, 2$ for arbitrary $s$ and for $k= 3$ with $s = 0$. We expect this to be true in general.
This is seemingly
related to the metaplectic correction in geometric quantization, something
we plan to address in more detail in a future publication.

Secondly, if we consider $2n$-manifolds with the full
$SO(2 n)$ holonomy, we do not expect purely gravitational anomalies
except for
$2 n = 4 k +2$, $k =0, \, 1,\, 2$, etc.
This is because the index density from
which the anomaly is descended, namely,
$\Tr R^{n + 1}$ vanishes
by virtue of the antisymmetry of $R $ as an element of the
algebra of $SO( 2 n)$.
In our case, we consider the restriction to
holonomies in $U(k) \subset SO(2 k)$, so we do not have the
transformations which can combine the 
$\underline{U(k)}$-valued curvatures
into a real antisymmetric matrix in $\underline{SO(2 k)}$.

The existence of the purely gravitational contributions
is related to the fact that the Dolbeault index is nonzero for
even dimensions, in a way similar 
to the argument given
in \cite{ag-witten} for fermions. 
For the gravitational anomaly for fermions in a general dimension $2n$, one can consider
the compactification of the manifold as $M_2 \times M_{2n-2}$, where
$M_2$ is two-dimensional and $M_{2n -2}$ is taken to be compact.
One can then consider the anomaly for Lorentz transformations (or diffeomorphisms)
on $M_2$. 
The effect of the remaining $(2n -2)$ dimensions is a multiplicative factor
corresponding to the number of zero modes of the relevant kinetic operator, i.e.,
the Dirac operator,
on $M_{2n -2}$. The anomaly in two dimensions, namely on $M_2$, 
then implies a nonzero anomaly on $M_{2n}$
 if the Dirac operator has a nonzero index on $M_{2n -2}$.
 This is possible for fermions only if $2 n -2 = 4 k$. 
 This reasoning works because
the anomaly may be viewed as a short distance effect arising from issues of regularization
and hence the compactification does not affect the final answer.
 For the case of interest to us, the Dolbeault operator has a nonzero index
 generically for any even dimension, in particular on $M_{2 k -2}$.
 Thus we should expect a gravitational anomaly with the
Dolbeault index 
density in $2k +2$ dimensions as the starting point for the descent procedure.

However, we may note that, although we do have a nonzero gravitational
contribution, there is a remnant in the final expressions from
the vanishing of $\Tr R^{n+1}$
due to the
antisymmetry property of $R $ (if it has values in $\underline{SO(2k}$).
Once we have combined $A$ with $\omega^0$ as in
$A+ ( s+ {\half } k ) \omega^0$, 
there is a left-over purely gravitational piece
in some cases.
In $(2+1)$ dimensions, this is given by
the last term in the braces in (\ref{3d}).
This has been interpreted as what is needed to cancel
the gravitational anomaly due to the chiral field on the edge in the case of a finite
droplet. In $(2+1)$ dimensions, the chiral field
on the edge lives in 1+1 dimensions, and produces an anomaly for
the Lorentz connection $\omega$.
For the $(4+1)$-dimensional case, 
the edge field is in $3+1$ dimensions.
The gravitational fields are valued in $\underline{SO(3,1)}$ or
$\underline{SO(4)}$ after a Euclidean continuation.
A chiral field would couple to one of the chiral components
in the splitting $SO(4) \sim SO(3)_L \otimes SO(3)_R$.
In this case, there is no Lorentz anomaly by the same reasoning as 
related to the antisymmetry of $R$ with values in the algebra of the orthogonal group.
Thus we should expect no purely non-Abelian
gravitational part in the action.
This is in agreement with what we find
in (\ref{CP22}), where there is no purely non-Abelian
gravitational term.

\section{Discussion}
In this paper, we have given a general expression
(\ref{24d}) for the topological part of the bulk effective action
for quantum Hall systems
in arbitrary even spatial dimensions. Explicit detailed formulae for the action are given in (2+1), (4+1) and (6+1) dimensions. 
The background metric and gauge field can be arbitrary in the sense that
fluctuations of the metric and the gauge field around a given background, but
which do not
change the topological class of the background,
are included. This action thus yields the topological terms in the response of the system (or correlation functions of the source currents) to changes in the gauge and gravitational fields.
The terms which involve only the gauge field had been obtained
earlier for the lowest Landau level in a large $N$ simplification, where $N$ denotes the
degeneracy of the Landau level \cite{DK}- \cite{KN5}.
Terms which involve both the gauge and the gravitational fields
provide a generalization of the well known Wen-Zee term in
the $(2+1)$ dimensional case.
Since these are subdominant in $N$, they were not 
evident in the leading large $N$ calculations.
(Some metric-dependent subdominant terms, including some
gauge-gravity mixing terms, were already in 
\cite{DK}-\cite{KN5}, but they were not explicitly stated in terms of the
curvatures, since a fixed gravitational background was used.)

The main justification for the effective action (\ref{24d})  is that the current 
densities obtained from it correctly reproduce
the degeneracies of the Landau levels via the Dolbeault index theorem.
In $(2+1)$ dimensions, our results agree with the
effective action which has been obtained by other authors by
different techniques.
The approach in \cite{KW2} uses a Dolbeault index density as well.
However, the starting point there is the index density in four dimensions.
A path in this space is considered as the time direction and a 
descent procedure from four dimensions to the $(2+1)$-dimensional world
of this line and the two-dimensional transverse space is used.
There are other important considerations in \cite{KW2}, including going
beyond the topological terms, but on questions for which our work has overlap with
this paper, the results agree.

More generally, for the effective action in
$(2k +1)$ dimensions, there are two index densities we can consider, in
$2k$ dimensions and in $(2k+2)$ dimensions.
The first one is relevant for the degeneracy and can be used to obtain
many of the terms in the effective action. However, as explained after (\ref{24}),
we may think of the action as also obtained via the descent procedure from
the Dolbeault index density in $(2k+2)$ dimensions.
The latter can be used to identify the purely gravitational terms related to gravitational anomalies and to clarify terms involving non-Abelian
gauge fields. It is useful to consider both index densities
together as they highlight  complementary aspects of the problem.

We have considered only fully filled Landau levels on manifolds without boundary.
The case of quantum Hall droplets, the action for the edge excitations which exist in such cases
and the interplay between the bulk and boundary actions are clearly the next set of interesting questions, to be taken up in future.
Also, beyond the milieu of exploring the geometry of
the quantum Hall effect in arbitrary dimensions,
geometry and topology, we may note that
quantum Hall effect in higher dimensions has been of interest
for spin Hall effect and for considerations on gravity.
The results of this paper may therefore be of specific interest in such contexts
as well.

\vskip .2in
This research was supported in part by the U.S.\ National Science
Foundation grants PHY-1417562, PHY-1519449
and by PSC-CUNY awards. 

\section*{Appendix}
\def\theequation{A\arabic{equation}}
\setcounter{equation}{0}

\subsection*{Basic features and geometry of ${\mathbb{CP}}^k$ spaces}

Let $t_A$ denote the generators of $SU(k+1)$ as matrices in the fundamental representation, normalized so that $\Tr (t_A t_B) = {1 \over 2} \delta_{AB}$. These generators are classified into three groups. The ones corresponding to the $SU(k)$ part of
$U(k) \subset SU(k+1)$ will be denoted by
$t_a$, $a =1, ~2, \cdots , ~ k^2 -1$ while the
generator for the $U(1)$
direction of the subgroup $U(k)$ will be denoted by 
$t_{k^2+2k}$. The $2k$ remaining generators of $SU(k+1)$ which are not in $U(k)$ are the coset generators, denoted by $t_\alpha$, $\alpha = k^2,\cdots, k^2+2k-1$. The coset generators can be further separated into the raising and lowering type $t_{\pm i}= t_{k^2+2i-2} \pm i t_{k^2+2i-1}, ~ i = 1, \cdots,k$. 

We can now use a $(k+1) \times (k+1)$ matrix $g$ in the fundamental representation of $SU(k+1)$ to parametrize ${\mathbb{CP}}^k$, by making the identification $g \sim gh$, where $h \in U(k)$. 
We can use the freedom of $h$ transformations to write $g$ as a function of the real coset coordinates $x^I$, $I=1,\cdots,2k$. The relation between the complex coordinates $z^i,~\bar{z}^i$ in (16) and $x^I$ is the usual one, 
\beq
z^i = x^{2i-1} + i x^{2i}~~~,~~~\bar{z}^i = x^{2i-1} - i x^{2i}~~~,~i=1,\cdots,k
\label{0A}
\eeq
We can write
\beq
g^{-1}dg = \big( -i E^{k^2+2k} t_{k^2+2k} -i E^{a} t_{a} -i E^{\alpha} t_{\alpha} \big)
\label{1A}
\eeq
$E^{\alpha}$ are 1-forms corresponding to the
frame fields in terms of which the Cartan-Killing metric on ${\mathbb{CP}}^k$ is given by
\beq
ds^2 = g_{ij} dx^i dx^j = E^\alpha_i E^\alpha_j dx^i dx^j
\label{2A}
\eeq
The K\"ahler two-form on ${\mathbb{CP}}^k$ is written as
\beqar
\Omega & = & -i \sqrt{{2k \over {k+1}}} \tr \left( t_{k^2 + 2k} ~ g^{-1}dg \wedge g^{-1}dg \right)
\nonumber
\\ & = & -{1 \over 4} \sqrt{{2k} \over {k+1}} f_{(k^2+2k)\alpha\beta} ~E^\alpha \wedge E^{\beta} ~=  -{ 1 \over 4} \epsilon_{\alpha\beta} ~E^\alpha \wedge E^{\beta}
\label{3A}
\eeqar
$f_{ABC}$ are the $SU(k+1)$ structure constants, where $[t_A,~t_B] = i f_{ABC} \,t_C$. In deriving the last line we used the fact that $f_{(k^2+2k)\alpha\beta}=  \sqrt{{k+1} \over {2k}}~ \epsilon_{\alpha\beta}$, where $\epsilon_{\alpha\beta} = 1$ if $\alpha= 2i-1, ~\beta= 2i, ~i=1, \cdots,k$. 

The K\"ahler two-form $\Omega$ can also be written in terms of the local complex coordinates in the more familiar form
\beq
\Omega = i \Bigr[ {{dz \cdot d\bar{z}} \over {1 + z\cdot \bar{z}}} -  {{\bar{z} \cdot dz ~z \cdot d\bar{z}} \over {(1 + z\cdot \bar{z})^2}}\Bigl]
\label{4A}
\eeq
where 
\beq
g_{i, k+1}= {z_i \over \sqrt{1+ z \cdot \bar{z}}} ~, i=1,\cdots, k ~~~~~~~~~~~~g_{k+1, k+1}= {1 \over \sqrt{1+ z \cdot \bar{z}}}
\label{5A}
\eeq
was used in (\ref{3A}).

The volume of ${\mathbb{CP}}^k$ is normalized so that
\beq
\int_{\mathbb{CP}^k} ~\Bigl({ \Omega \over {2 \pi}} \Bigr)^k = 1
\label{6A}
\eeq

The Maurer-Cartan identity along with (\ref{1A}) leads to
\beqar
dE^{k^2+2k} & = & -\half f^{(k^2+2k)\alpha \beta} E^{\alpha} \wedge E^{\beta} = 2 \sqrt{{k+1} \over {2k}} ~\Omega \nonumber \\
dE^a + \half f^{abc} E^b \wedge E^c & = & - \half f^{a \alpha \beta} E^{\alpha}\wedge E^{\beta}
\label{7A}\\
dE^{\alpha} & = & - f^{\alpha A  \beta} E^{A} \wedge E^{ \beta}
\nonumber
\eeqar
Combining the $2k$ frame fields $E^{\alpha}$ into holomorphic combinations and using (\ref{7A}) we can identify the spin connection for the complex cotangent space $T^*_cK$,
\beqar
{\cal E}^I & \equiv &  E^{2I-1} + i\, E^{2I}, \hskip .3in d{\cal E}^I + { \omega}_*^{IJ} {\cal E}^J =0 ~~,~~ I=1, \cdots, k \nonumber\\
{\omega}_* & = & -i \Bigl( \sqrt{{k+1} \over {2k}} E^{k^2+2k} (-\bold{1}) +  E^a (-t_a)^T \Bigr)
\label{8A}
\eeqar
where $\bold{1}$ is the $k \times k$ identity matrix and $t_a$ are the $SU(k)$ matrices in the fundamental representation and the superscript $T$ on $t_a$ indicates the transpose.
A basis for the tangent space $T_cK$ is given by
vector fields dual to ${\cal E}^I$. By differentiating
the relation ${\cal E}^I_a\, ({\cal E}^{-1})^b_I = \delta^b_a$, we can identify the
spin connection for $T_cK$ as
\beq
{\omega} =  -i \Bigl( \sqrt{{k+1} \over {2k}} E^{k^2+2k} \,\bold{1} +  E^a \,t_a  \Bigr)
\label{8Aa}
\eeq
Notice that
$ {\bold 1} \rightarrow (-{\bold 1})$ and $t_a \rightarrow (-t_a)^T$ appearing in (\ref{8A}) correspond to the conjugation
operation for the Lie algebra. Thus the Lie algebra
conjugation operation in going from the cotangent space
to the tangent space is exactly as expected.

Using (\ref{7A}) we can also derive the curvature two-form as
\beqar
R & = & d\omega + \omega \wedge \omega \nonumber \\
& =& -i \Bigl( {{k+1} \over k} \Omega ~\bold{1} - \half f^{a \alpha \beta} E^{\alpha} \wedge E^{\beta} ~t_{a} \Bigr) \label{9A}\\
& = & -i \Bigl( R^0 \bold{1} + R^a t_a \Bigr) \nonumber
\eeqar
where $R^0 = {{k+1} \over k} \Omega$ and $R^a = - \half f^{a \alpha \beta} E^{\alpha} \wedge E^{\beta}$. 

For $\mathbb{CP}^k$ spaces 
\beq
\int_{\mathbb{CP}^k} {\rm td}\, (T_c K) \big\arrowvert_{2k} = 1
\label{10A}
\eeq
where ${\rm td}\,(T_c K)$ is the Todd class in the complex tangent space and in 
(\ref{10A}) the $2k$-form is selected as the integrand. Explicitly,
the Todd class has the expansion given in (\ref{25}) as
\beq
{\rm td} = 1 + {1\over 2} \, c_1 +{1\over 12} ( c_1^2 + c_2) + {1\over 24} c_1\, c_2
+ {1\over 720} ( - c_4 + c_1 c_3 + 3 \, c_2^2 + 4\, c_1^2 \,c_2 - c_1^4) + \cdots
\label{11A}
\eeq
where $c_i$ are the Chern classes. The first few Chern classes can be easily evaluated using (\ref{26}) as
\beqar
c_1 & = & \Tr ~{iR \over 2\pi} = (k+1) { \Omega \over 2\pi} \nonumber \\
c_2 & = & \half \Bigl[ (\Tr {iR \over 2\pi})^2 - \Tr ({iR \over 2\pi})^2 \Bigr] = \half k (k+1) ({ \Omega \over 2\pi})^2 
\label{12A}
\eeqar
In deriving the expression for $c_2$ we used the fact that
\beqar
R^a \wedge R^a & = &\quarter f^{a \alpha\beta}f^{a \gamma\delta} E^{\alpha}E^{\beta}E^{\gamma}E^{\delta} = -2~{k+1 \over k}  \Omega^2 \nonumber \\
\Tr \bigl[{{iR\wedge iR}} \bigr]& = & k (R^0)^2 + \half (R^a)^2 = (k+1) \Omega^2
\label{13A}
\eeqar
More generally the Chern classes for ${\mathbb{CP}}^k$ can be written as
\beq
c_i =  {{k!} \over {i! (k-i)!}} \left( {\Omega \over 2\pi}\right) ^i
\label{13AA}
\eeq
Using (\ref{6A}) and (\ref{12A}), we can easily check the validity of (\ref{10A}) for $\mathbb{CP}^1$, 
$\mathbb{CP}^2$ and $\mathbb{CP}^3$, the needed integrals being
\beqar
\int_{\mathbb{CP}^1} c_1 & = & 2 \int { \Omega \over 2\pi} =2 \nonumber \\
\int_{\mathbb{CP}^2} c_1^2 + c_2 & = & (3^2+3) \int \left({ \Omega \over 2\pi}\right)^2 =12 \nonumber \\
\int_{\mathbb{CP}^3} c_1 c_2 & = & 4 \times 6 \int \left({ \Omega \over 2\pi}\right)^3 =24 
\label{14A}
\eeqar

In formulating QHE on ${\mathbb{CP}}^k$, we choose $U(1)$ and $SU(k)$ background gauge fields proportional to $E^{k^2+2k}_i$ and $E^a_i$. In particular 
\beqar
A^{k^2+2k} & = &- i n \sqrt{{{2k} \over {k+1}}} \tr (t_{k^2+2k} g^{-1} dg ) = {n \over 2} \sqrt{{{2k} \over {k+1}}} E^{k^2+2k} \nonumber \\
A^a = E^a & = & 2i \Tr (t^a g^{-1} dg) 
\label {Abar}
\eeqar
The corresponding $U(1)$ and $SU(k)$ background field strengths are
\beqar
F & = & n \Omega = - {n \over 4} \sqrt{{{2k} \over {k+1}}} f^{(k^2+2k)\alpha\beta} E^{\alpha}\wedge E^{\beta}  \nonumber \\
F^{a} & = & - f^{a\alpha\beta} E^{\alpha} \wedge E^{\beta}
\label{barF}
\eeqar
Notice that $F^a$ in (\ref{barF}) does not depend on $n$, while the Abelian field is proportional to $n$. 
We see from (\ref{barF}) that the background field strengths are constant in the appropriate frame basis, proportional to the $U(k)$ structure constants. It is in this sense that the field strengths in (\ref{barF}) correspond to uniform magnetic fields appropriate in defining QHE.




\begin{thebibliography}{99}

\bibitem{books} See for example:
R.E. Prange and S.M. Girvin, {\it The Quantum Hall Effect},
2nd ed. (Springer-Verlag, Berlin, 2012); Z.F. Ezawa,
{\it Quantum Hall Effects} (World Scientific, Singapore, 2008).


\bibitem{viscosity} J.E. Avron, R. Seiler and P.G. Zograf, {\it Viscosity of quantum Hall fluids}, \PRL { \bf 75}, 697 (1995).

\bibitem{Read} N. Read, {\it Non-Abelian adiabatic statistics and Hall viscosity in quantum Hall states and $p_x + i p_y$ paired superfluids}, \PR {\bf B79}, 045308 (2009); N. Read and E.H. Rezayi, {\it Hall viscosity, orbital spin, and geometry: paired superfluids and quantum Hall systems}, \PR { \bf B84}, 085316 (2011).

\bibitem{haldane} F.D.M. Haldane, {\it Fractional Quantization of the Hall Effect: A Hierarchy of Incompressible Quantum Fluid States},
Phys. Rev. Lett. 51, 605 (1983)
\PRL ~{\bf 51},
605 (1983).

\bibitem{haldane2}
F. D. M. Haldane and E. H. Rezayi,
{\it Periodic Laughlin-Jastrow wave functions for the fractional quantized Hall effect},
Phys. Rev. B 31, 2529(R) (1985).

\bibitem{frohlich} J. Fr\"ohlich and U.M. Studer, {\it $U(1)\times SU(2)$ gauge invariance of nonrelativistic quantum mechanics and generalized Hall effects},
Commun. Math. Phys. {\bf 148}, 553 (1992);
{\it Gauge invariance and current algebra in nonrelativistic many body theory}, \RMP { \bf 65},
733 (1993).

\bibitem{WZ} X. Wen and A. Zee, { \it Shift and spin vector: New topological quantum numbers for the Hall fluids}, \PRL { \bf 69}, 953 (1992).


\bibitem{HS} C. Hoyos and D.T. Son, {\it Hall viscosity and electromagnetic response}, \PRL {\bf 108}, 066805 (2012); D.T. Son, {\it Newton-Cartan geometry and the quantum Hall effect},  arXiv:1306.0638.


\bibitem{AG1} A.G Abanov and A. Gromov, {\it Electromagnetic and gravitational responses of two-dimensional non-interacting electrons in background magnetic field}, \PR  { \bf B90}, 014435 (2014); A. Gromov and A. Abanov, { \it Density curvature response and gravitational anomaly}, \PRL { \bf 113}, 266802 (2014).

\bibitem{AG2} A. Gromov, G. Cho, Y. You, A.G. Abanov and E. Fradkin, {\it Framing anomaly in the effective theory of the fractional quantum Hall effect}, { \PRL} { \bf 114}, 016805 (2015).

\bibitem{Wiegmann} T. Can, M. Laskin and P. Wiegmann, {\it Fractional Quantum Hall Effect in a Curved Space: Gravitational Anomaly and Electromagnetic Response}, \PRL { \bf 113}, 046803 (2014); {\it Geometry of Quantum Hall States: Gravitational Anomaly and Kinetic Coefficients}, Ann. Phys. { \bf 362} 752 (2015).



\bibitem{KW1} S. Klevtsov and P. Wiegmann, {\it Geometric adiabatic transport in quantum Hall states}, \PRL { \bf 115} 086801 (2015).

\bibitem{BR} B. Bradlyn and N. Read, {\it Topological central charge from Berry curvature: gravitational anomalies in trial wavefunctions for topological phases}, \PR { \bf B91}, 165306 (2015).

\bibitem{KW2} S. Klevtsov, X. Ma, G. Marinescu and P. Wiegmann, {\it Quantum Hall effect, Quillen metric and holomorphic anomaly}, , arXiv:1510.06720.

\bibitem{CR} A. Cappelli and E. Randellini, {\it Multipole expansion in the quantum Hall effect}, arXiv:1512.02147.


\bibitem{HZ} S.C. Zhang and J.P. Hu, {\it A four dimensional generalization of the quantum Hall effect}, {Science} {\bf 294} (2001) 823; 
J.P. Hu and S.C. Zhang, {\it Collective excitations at the boundary of a 4D quantum Hall droplet}, cond-mat/0112432.

\bibitem{KN1} D. Karabali and V.P. Nair, {\it Quantum Hall effect in higher dimensions}, \NP { \bf B641}, 533 (2002).

\bibitem{KN2} D. Karabali and V.P. Nair, {\it The effective action for edge states in higher dimensional quantum Hall systems}, \NP { \bf B679}, 427 (2004).

\bibitem{KN3} D. Karabali and V.P. Nair, {\it Edge states for quantum Hall droplets in higher dimensions and a generalized WZW model}, \NP { \bf 697}, 513 (2004).

\bibitem{KN4} D. Karabali, V.P. Nair and S. Randjbar-Daemi,  {\it Fuzzy spaces, the M(atrix) model and quantum Hall effect}, published in  {\it From fields to strings: Circumnavigating theoretical physics}, edited by M. Shifman et al., vol. 1, 831-875 (World Scientific, Singapore, 2005).

\bibitem{everyone} M. Fabinger, {\it Higher-dimensional quantum Hall effect in string theory}, { JHEP} {\bf 0205}, 037 (2002); 
Y.X. Chen, B.Y. Hou, B.Y. Hou, {\it Non-commutative algebra of functions of 4-dimensional quantum Hall droplet},{ Nucl. Phys.} {\bf B638}, 220 (2002); Y. Kimura, {\it Noncommutative gauge theory on fuzzy four-sphere and matrix model}, { Nucl. Phys.} {\bf B637}, 177 (2002); H. Elvang, J. Polchinski, {\it The quantum Hall effect on} $R^4$, hep-th/0209104; B.A. Bernevig, C.H. Chern, J.P. Hu, N. Toumbas, S.C. Zhang, {\it Effective field theory description of the higher dimensional quantum Hall liquid}, 
{ Ann. Phys.} {\bf 300}, 185 (2002);
B. A. Bernevig, J.P. Hu, N. Toumbas, S.C. Zhang, {\it Eight-dimensional quantum Hall effect and ``octonions"}, { Phys. Rev. Lett.}
{\bf 91}, 236803 (2003); 
 G. Meng, {\it Geometric construction of the quantum Hall effect in all even dimensions}, { J. Phys.} {\bf A36}, 9415 (2003);  V.P. Nair and S. Randjbar-Daemi, {\it Quantum Hall effect on} $S^3$, {\it edge states and fuzzy} $S^3 / Z_2$, { Nucl. Phys.}
{\bf B679}, 447 (2004); A. Jellal, {\it Quantum Hall effect on higher-dimensional spaces}, { Nucl. Phys.} {\bf B725}, 554 (2005).
  
\bibitem{dolan} B. Dolan, {\it The spectrum of the Dirac operator on coset spaces with homogeneous gauge fields}, { JHEP} {\bf 0305}, 18 (2003);
K. Hasebe, {\it Higher dimensional quantum Hall effect as A-class topological insulator}, { Nucl. Phys.} {\bf B886}, 952 (2014).

  
\bibitem{poly1} A.P. Polychronakos, {\it Chiral actions from phase space (quantum Hall) droplets}, {Nucl. Phys.} {\bf B705}, 457 (2005); {\it Kac-Moody theories for colored phase space (quantum Hall) droplets}, { Nucl. Phys.} {\bf B711}, 505 (2005).


\bibitem{DK} D. Karabali, {\it Electromagnetic interactions of higher dimensional quantum Hall droplets}, \NP { \bf B726}, 407 (2005); {\it Bosonization of the lowest Landau level in arbitrary dimensions: Edge and bulk dynamics}, \NP { \bf B750}, 265 (2006).

\bibitem{VPN} V.P. Nair, {\it The matrix Chern-Simons one-form as a universal Chern-Simons theory}, \NP  { \bf B750}, 289 (2006).

\bibitem{KN5} D. Karabali and V.P. Nair, {\it Quantum Hall effect in higher dimensions, matrix models and fuzzy geometry}, {Journal of Physics} {\bf A39}, 12735 (2006).

\bibitem{wen2d} X.G. Wen, {\it Chiral Luttinger liquid and the dge excitations in the fractional quantum Hall states}, \PR~ {\bf B41}, 12838 (1990); D.H. Lee and W.G. Wen, {\it Edge excitations in the fractional quantum Hall liquids}, {Phys. Rev. Lett.} {\bf 66}, 1765 (1991); M. Stone, {\it Schur functions, chiral bosons, and the quantum Hall effect edge states}, \PR~ {\bf B42}, 8399 (1990); {\it Edge waves in the quantum Hall effect}, {Ann. Phys.} (NY) {\bf 207},  38 (1991); J. Fr\"ohlich and T. Kepler,  {\it Universality in quantum Hall systems}, \NP~ {\bf B354}, 369 (1991).

\bibitem{CalHarv} C.G. Callan and J.A. Harvey, {\it Anomalies and fermion zero modes on strings and domain walls}, Nucl. Phys. {\bf B250}, 427 (1985).

\bibitem{eguchi} T. Eguchi, P.B. Gilkey and A.J. Hanson, {\it Gravitation, gauge theories and differential geometry}, Phys. Rep. { \bf 66}, 213 (1980).

\bibitem{anom}
R.A. Bertlmann, {\it Anomalies in Quantum Field Theory} (Oxford University Press, New York, 1996);
S. Treiman, R. Jackiw, B. Zumino and E. Witten,
{\it Current Algebra and Anomalies}  (World Scientific, Singapore, 1985).

\bibitem{ag-witten} L. Alvarez-Gaum\'e and E. Witten, {\it Gravitational Anomalies},
\NP~{\bf B234}, 269 (1984).


\end{thebibliography}
\end{document}